\documentclass[]{spie}  

\usepackage{amsmath,amsfonts,amssymb}
\usepackage{graphicx}
\usepackage[colorlinks=true, allcolors=blue]{hyperref}

\newcommand {\micron}{\mbox{$\mu$m}}
\newcommand\arcsec{$^{\prime\prime}$}

\title{Low Wind Effect on VLT/SPHERE : impact, mitigation strategy, and results}

\author[a]{J. Milli}
\author[c]{M. Kasper}
\author[a]{P. Bourget}
\author[a]{C. Pannetier}
\author[b]{D. Mouillet}
\author[d,e]{J.-F. Sauvage}
\author[a]{C. Reyes}
\author[d,e]{T. Fusco}
\author[f]{F. Cantalloube}
\author[a]{K. Tristram}
\author[a]{Z. Wahhaj}
\author[c]{J.-L. Beuzit}
\author[g]{J. H. Girard}
\author[h,i]{D. Mawet}
\author[j]{A. Telle}
\author[d]{A. Vigan}
\author[k]{M. N'Diaye}

\affil[a]{European Southern Observatory (ESO), Alonso de C\'ordova 3107, Vitacura, Casilla 19001, Santiago, Chile}
\affil[b]{Universit\'e Grenoble Alpes, IPAG, F-38000 Grenoble, France }
\affil[c]{ESO, Karl-Schwarzschild-Stra{\ss}e 2, 85748 Garching, Germany}
\affil[d]{Aix Marseille Universit\'e, CNRS, LAM (Laboratoire d'Astrophysique de Marseille) UMR 7326, 13388, Marseille, France}
\affil[e]{ONERA, The French Aerospace Lab, BP72, 29 avenue de la Division Leclerc, 92322 Chatillon Cedex, France}
\affil[f]{Max Planck Institute for Astronomy, K\"onigstuhl 17, D-69117 Heidelberg, Germany}
\affil[g]{Space Telescope Science Institute, 3700 San Martin Drive, Baltimore, MD 21218, USA}
\affil[h]{Department of Astronomy, California Institute of Technology, 1200 E. California Blvd, MC 249-17, Pasadena, CA 91125 USA}
\affil[i]{Jet Propulsion Laboratory, California Institute of Technology, 4800 Oak Grove Drive, Pasadena, CA 91109, USA}
\affil[j]{ACM Coatings GmbH, Rudelsburgpromenade 20c, 06628 Naumburg - Bad K\"osen, Germany}
\affil[k]{Universit\'e C\^ote d'Azur, Observatoire de la C\^ote d'Azur, CNRS, Laboratoire Lagrange, Bd de l'Observatoire, CS 34229, 06304 Nice cedex 4, France}

\authorinfo{Further author information: Julien Milli: E-mail: jmilli@eso.org}

\begin{document} 
\maketitle

\begin{abstract}

The low wind effect is a phenomenon disturbing the phase of the wavefront in the pupil of a large telescope obstructed by spiders, in the absence of wind. It can be explained by the radiative cooling of the spiders, creating air temperature inhomogeneities across the pupil. Because it is unseen by traditional adaptive optics (AO) systems, thus uncorrected, it significantly degrades the quality of AO-corrected images. We provide a statistical analysis of the strength of this effect as seen by VLT/SPHERE after 4 years of operations. We analyse its dependence upon the wind and temperature conditions. We describe the mitigation strategy implemented in 2017: a specific coating with low thermal emissivity in the mid-infrared was applied on the spiders of Unit Telescope 3. We quantify the improvement in terms of image quality, contrast and wave front error using both focal plane images and measured phase maps.

\end{abstract}


\section{INTRODUCTION}
\label{sec_intro}  

SPHERE (Spectro-Polarimetric High-contrast Exoplanet REsearch\cite{Beuzit2008}) is an instrument dedicated to giant planet searches using direct imaging at the Very Large Telescope (VLT). For bright stars (R magnitude below 6) under fair seeing and good coherence time, it routinely achieves 90\% Strehl at H-band, with a raw contrast of $10^{-5}$ at 500\,mas\cite{Milli2017_SPHERE, Mouillet2018} . To do that, it is fed by an extreme Adaptive Optics (AO) system called SAXO \cite{Fusco2016,Sauvage2016_SAXO} , which operates at a frequency up to 1.38 kHz on bright targets with a 40x40 spatially filtered Shack-Hartmann (SH) wavefront sensor (WFS) and a 41x41 piezoelectric high-order deformable mirror. This high level of performance revealed limitations coming from the telescope environment, that were hardly detectable with the previous generation of AO systems on the VLT such as NAOS\cite{Rousset2003} on NACO or the MACAO systems \cite{Arsenault2003} used on SINFONI, CRIRES or with the VLT-interferometer. Two main effects were identified early during the commissioning of SPHERE in May 2014. First, spurious vibrations introduced by the secondary mirror were detected and removed by switching off the 50 Hz field stabilisation when SPHERE is used. Secondly, a degradation of the performance of the instrument was more problematic to diagnose and to mitigate, and it took almost three years to find and implement a satisfactory solution.  As the effect occurs exclusively when the wind speed near the ground is very low, it was coined the Low-Wind Effect (LWE)\cite{Sauvage2015,Sauvage2016} . 

\subsection{Description of the effect and data used for the analysis}

   \begin{figure} [b]
   \begin{center}
   \begin{tabular}{c} 
   \includegraphics[width=0.8\hsize]{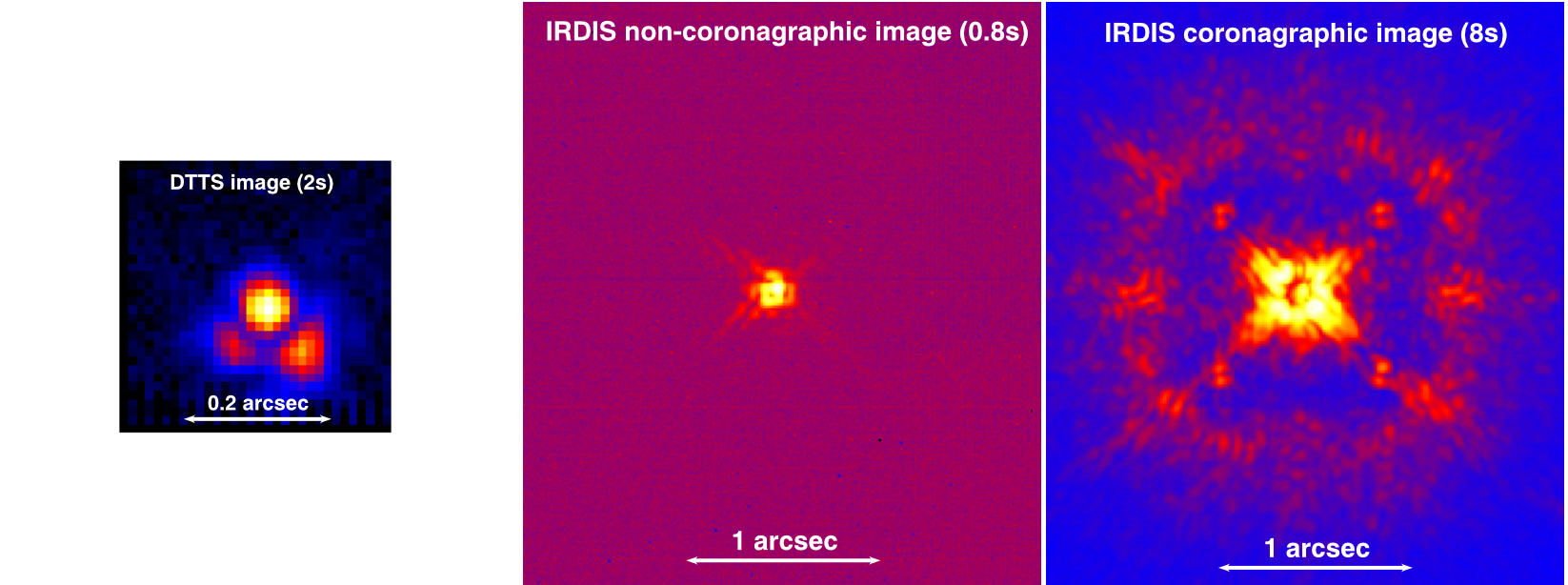}
	\end{tabular}
	\end{center}
   \caption[ex] 
   { 
   Images in the H-band obtained during a LWE event at a few minutes interval on a bright star. Left: DTTS image. Middle: IRDIS non-coronagraphic image with the shortest possible exposure time in full frame. Right: deeper IRDIS coronagraphic exposure, while the waffle pattern is applied on the DM to create four PSF replica. 
   \label{fig_comp_DTTS_coro_nocoro} 
   }
   \end{figure} 

\begin{figure} [ht]
   \begin{center}
   \begin{tabular}{c} 
   \includegraphics[width=\hsize]{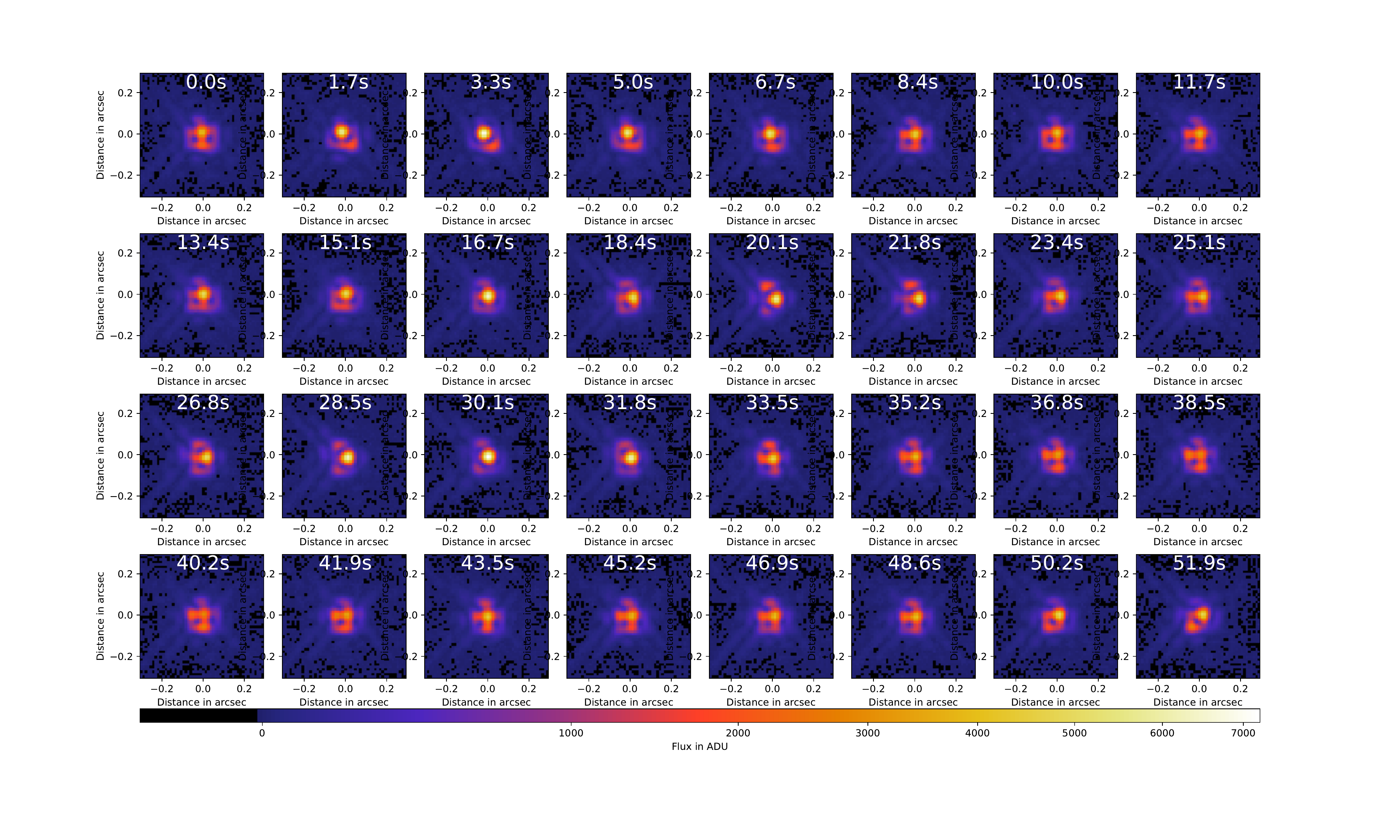}
	\end{tabular}
	\end{center}
   \caption[ex] 
   { 
   IRDIS images in the H2 filter, of 0.84s exposure time (minimum DIT in full frame mode), obtained during a strong LWE event (strength between 20\% and 50\%, wind speed of about 1.5 m/s). The number in each frame indicates the time elapsed from the beginning of the sequence. The colour range scales with the square root of the intensity.
   \label{fig_irdis_nocoro_sequence} 
   }
   \end{figure}

\begin{figure} [ht]
   \begin{center}
   \begin{tabular}{c} 
   \includegraphics[width=0.7\hsize]{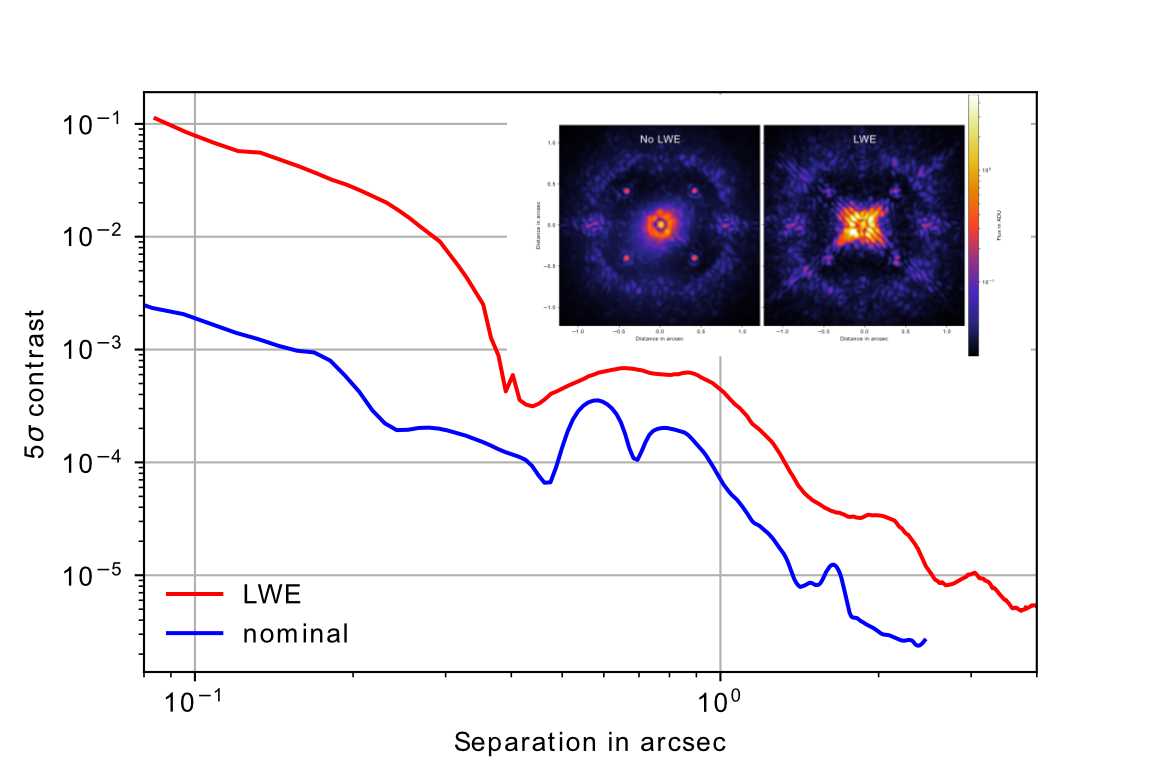}
	\end{tabular}
	\end{center}
   \caption[ex] 
   { 
   Comparison of two raw $5\sigma$ contrast curves obtained in similarly good seeing and coherence time conditions, but one with low wind speed (red curve, right image) and one with a mild 5m/s wind (blue curve, left image). The two images are at H band, with the waffle pattern imprinted on the deformable mirror, hence the bump at about 0.6\arcsec.
   \label{fig_irdis_comparison_with_without_LWE} 
   }
   \end{figure}

For mild to moderate LWE, non-coronagraphic focal planes images show bright secondary lobes at the location of the first Airy ring as shown on Fig. \ref{fig_comp_DTTS_coro_nocoro} (left and middle image). The middle image was taken with the near-infrared science camera IRDIS\cite{Dohlen2008} , while the left image was obtained with the Differential Tip-Tilt Sensor\cite{Baudoz2010} (DTTS), part of SAXO that is used for low-order near-infrared wavefront-sensing close to the coronagraph focus. An average image from this camera is saved every two seconds along with other telemetry parameters\footnote{for details on how to download the SAXO telemetry data, see Milli et al. 2017\cite{Milli2017_SPHERE} or the SPHERE User Manual}, and a large database of such images is now available. For the analysis presented here, we retrieved all the 576\,430 DTTS images collected between January, 1 2015 and May, 20 2018 when the DTTS loop was closed (e.g. excluding the 12\% of the time where the Zimpol sub-system was used, as  Zimpol is an optical imager that does not make use of this near-infrared sensor).  The DTTS uses 2\% of the light going to to the near-infrared arm and images the AO target star in the H-band in a window of $32\times32$ pixels corresponding to 0.35\arcsec$\times$0.35\arcsec. The signal-to-noise (S/N) can be low  for targets fainter than 10 at H, therefore an additional selection based on the S/N was performed and described in section \ref{subsec_LWE_strength}. These images are however available in much larger quantities compared to IRDIS non-coronagraphic images (middle image) that are typically taken only for flux calibration of coronagraphic frames. 
In Fig. \ref{fig_comp_DTTS_coro_nocoro}, the LWE appears as two secondary lobes but the number of lobes can vary from one to four in extreme cases. We show in Fig. \ref{fig_irdis_nocoro_sequence} a much stronger event with two to four secondary lobes. In a coronagraphic image, the LWE ruins the contrast at short separations as visible on Fig.  \ref{fig_comp_DTTS_coro_nocoro} right. Bright starlight residuals are visible near the coronagraph edge. As shown in Fig. \ref{fig_irdis_comparison_with_without_LWE}, the raw contrast curve is severely degraded by a factor about 50 at 0.1\arcsec{} and can be impacted up to several arcseconds. 

The typical timescale of evolution of the effect is one to two seconds. Fig. \ref{fig_irdis_nocoro_sequence} shows a typical evolution of the effect over one minute with the minimum Detector Integration Time (DIT) of 0.84s of IRDIS, corresponding to an effective frame rate of one image every 1.7s including readout. The similarity between an image and the next one suggests that the phase aberration typically evolves over a few seconds, but no higher frame rate images were obtained to confirm this property. 

\begin{figure} [ht]
   \begin{center}
   \begin{tabular}{c} 
   \includegraphics[width=0.8\hsize]{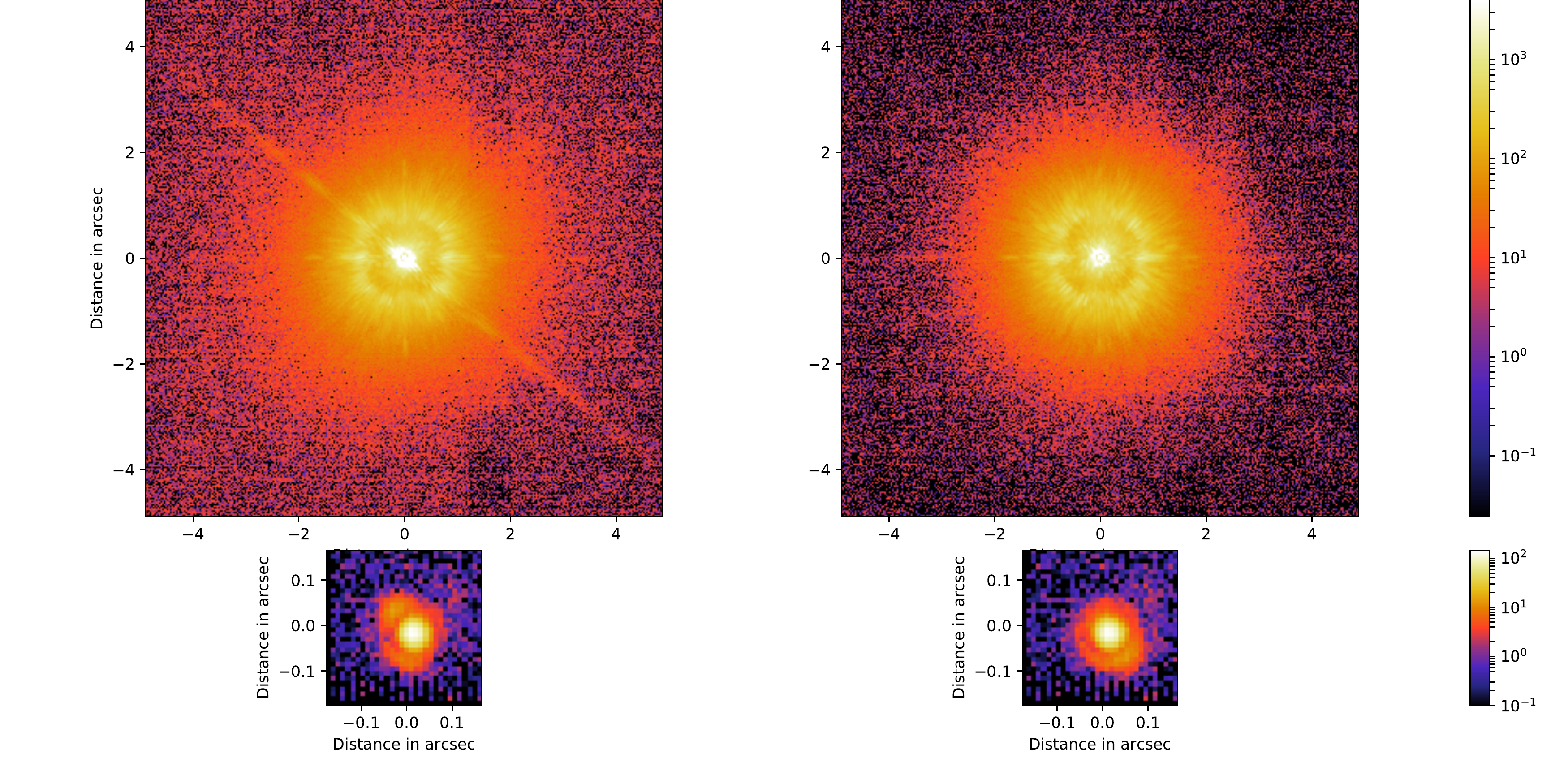}
	\end{tabular}
	\end{center}
   \caption[ex] 
   { Comparison during an observation with LWE between an image affected by the distorsion (left) and only mildly affected (right, taken one minute apart) due to the variability of the phenomenon. The colour scale is logarithmic to  enhance the diffraction spike at $45^\circ$. The top images are the coronagraphic H band images with IRDIS, the bottom images are the corresponding DTTS H band non-coronagraphic images. This data are part of a sequence on the debris disk host star 49\. Ceti\cite{Choquet2017} . A video of the sequence is available online at \url{https://doi.org/10.1117/12.2311499} (file video\_corono+DTTS.mov in the supplemental content section).
   \label{fig_spider_diffraction_spike} 
   }
   \end{figure}

Another remarkable feature is the presence of the spider diffraction spikes visible well outside the AO correction radius, as visible in Fig. \ref{fig_spider_diffraction_spike}. In pupil-stabilised coronagraphic images, the diffraction pattern of the spiders is indeed masked by the Lyot stop. However, the secondary lobes created in case of LWE behave like an off-axis source, hence the fact that the spiders can become visible. Because of the variability of the effect, the diffraction spikes can appear and disappear alternatively during an observing sequence, depending on the strength of the effect. This effect was already observed and understood well before the advent of adaptive optics by Couder using an 80cm telescope\cite{Couder1949} . He could detect the effect visually by looking at the diffraction spikes of bright stars and confirm its origin by measuring a phase jump of 280nm in the pupil at the location of the spiders using the Foucault knife-edge effect. 

\subsection{Interpretation}

   \begin{figure} [ht]
   \begin{center}
   \begin{tabular}{c} 
   \includegraphics[width=\hsize]{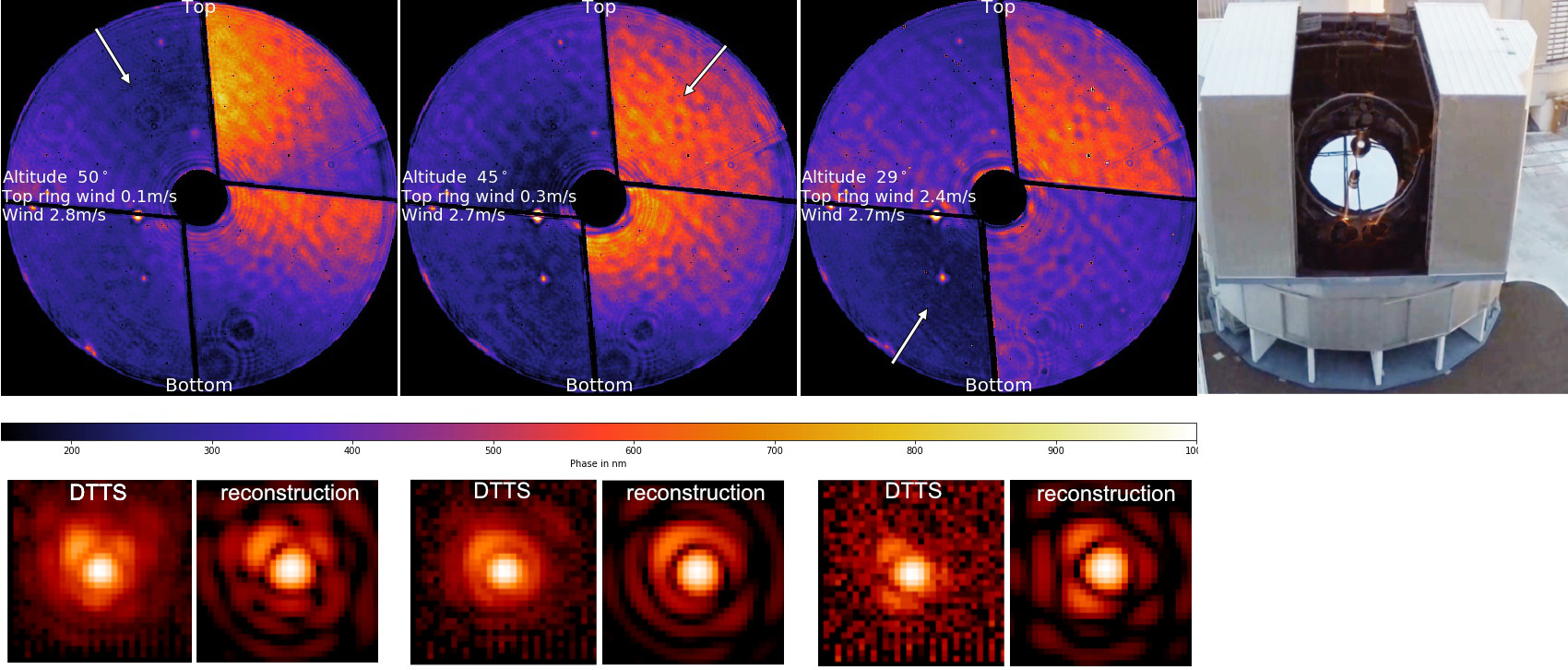}
	\end{tabular}
	\end{center}
   \caption[ex] 
   { 
 Top: Phase maps captured with the Zernike phase mask during the LWE night of October 8 2014 at three different telescope pointings. The white arrow indicates the direction of the wind projected on the pupil. The top of the image corresponds to the top of the dome (protecting from the wind as seen on the right image), whereas the bottom of the image is the closest to the main mirror doors. Bottom: corresponding DTTS and synthetic image showing that the combination of differential piston, tip and tilt seen in the pupil does indeed reproduce the observed PSF. Right: aerial picture from G. Huedepohl.
   \label{fig_zelda_phase_winddir} 
   }
   \end{figure} 

The Shack-Hartmann (SH) wavefront sensor is not sensitive to the LWE because it can only measure the slope of the wavefront, but SPHERE is equipped with a Zernike phase mask called Zelda\cite{Ndiaye2016} that can measure directly the phase in a given linearity range. We show in Fig. \ref{fig_zelda_phase_winddir} the phase in the pupil as measured by this sensor (top) and the corresponding PSF as measured by the DTTS. The phase map was calibrated in nm in order to reproduce the PSF morphology (cf reconstructed DTTS image at the bottom). A discontinuity of up to 500nm is visible between one quadrant of the pupil and the next one. In addition to this differential piston between quadrants, a differential tip/tilt is also present and necessary to explain the PSF morphology.  

The analysis of the shape of the deformable mirror shows no anomaly similar to what is seen in the pupil, indicating that the deformable mirror is not responsible for introducing the phase aberration. A map of the variance of the  SH slopes shows stronger values around the spiders, indicating that the WFS is at least partially sensitive to the effect\cite{Sauvage2015} .

The origin of this phase error in the pupil is now well understood. Couder proposed in 1949\cite{Couder1949} that the spiders loose heat by radiation and cool down. The telescope spiders on the four Unit Telescopes (UT) of the VLT are equipped with thermocouples and we verified empirically that they are indeed between $1^\circ$ and $3^\circ$ colder than the ambient air during the night (see for instance Fig. \ref{fig_LWE_vs_windspeed_before_coating} middle, later). 

When the wind is low, there is little convection in the dome and the colder air at the contact of the spiders is not well mixed and can contaminate a significant area of each quadrant of the pupil. A one-dimensional Computer Fluid Dynamics simulation\cite{Sauvage2016}  showed that a discontinuity of $1^\circ$ downstream of a spider $3^\circ$ colder than the air can be obtained with a 0.3m/s wind speed in the pupil. This in turn can create an Optical Path Difference (OPD) of 100nm. To understand the 3D flow that settles in the pupil, we analysed the wind speed and direction for three different telescope pointings and show the result in Fig. \ref{fig_zelda_phase_winddir}. No general conclusions can be drawn. When the telescope is pointing down the wind (left and middle image with the white arrow coming from the top), the wind speed at the top ring (supporting the spiders) is almost zero and moving the telescope altitude axis or azimuth axis does not change much the phase. There seems to be no relation between the wind direction and the quadrant most affected by the differential piston. When the telescope is facing the wind (right image with the white arrow coming from the bottom), the wind speed at the top ring reaches almost the ambient wind speed and the quadrant the most affected by the differential piston is opposite the wind direction. This conclusion might not be generalisable to all LWE occurrences, but given the very few number of Zelda images obtained in LWE conditions, nothing else can be said for now. 

\section{Quantitive assessment of the LWE}

\subsection{The low-wind effect strength $\mathcal{S}$}
\label{subsec_LWE_strength}

\begin{figure} [ht]
   \begin{center}
   \begin{tabular}{c} 
   \includegraphics[width=0.3\hsize]{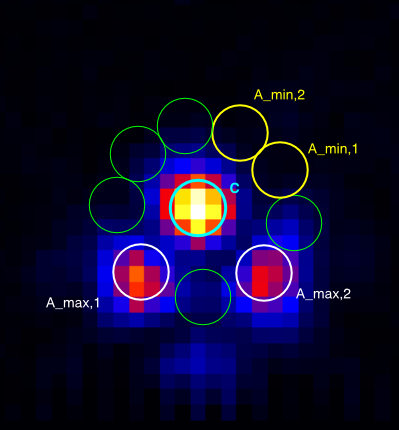}
	\end{tabular}
	\end{center}
   \caption[ex] 
   { 
The LWE strength $\mathcal{S}$ is defined as $\mathcal{S}=\frac{(A_{max,1}+A_{max,2})-(A_{min,1}+A_{min,2})}{C}$
   \label{fig_LWE_strength_definition} 
   }
   \end{figure}

In order to understand the conditions in which the LWE occurs, we derived a parameter, called the LWE strength $\mathcal{S}$, to estimate the strength of the LWE from the DTTS images. This parameter was defined as the ratio between the asymmetry $A$ on the first Airy ring and the central core of the PSF noted $C$: $\mathcal{S}= A/C$. $C$ was evaluated as the flux within a circular aperture of diameter $1\lambda/D$ centred on the best fit position of a Gaussian to the DTTS image. To assess A, we distributed identical apertures along the first Airy ring, and kept the 2 non-overlapping apertures of maximum flux, called $A_{max,1}$ and $A_{max,2}$, and the 2 non-overlapping apertures of minimum flux, called $A_{min,1}$ and $A_{min,2}$. We defined $A$ as $A=(A_{max,1}+A_{max,2})-(A_{min,1}+A_{min,2})$. The reason why 2 minimums and 2 maximums are used and not simply $A_{max,1}-A_{min,1}$  is because the former yields a higher LWE strength for a PSF with 2 lobes than for a PSF with only 1 lobe. An illustration of the aperture locations is given in Fig. \ref{fig_LWE_strength_definition}. The LWE strength directly gives an estimate of the amount of asymmetry in the first Airy ring as a fraction of the flux in the core.  

\subsection{Calibration of the LWE strength $\mathcal{S}$}

   \begin{figure} [ht]
   \begin{center}
   \begin{tabular}{c} 
   \includegraphics[width=0.8\hsize]{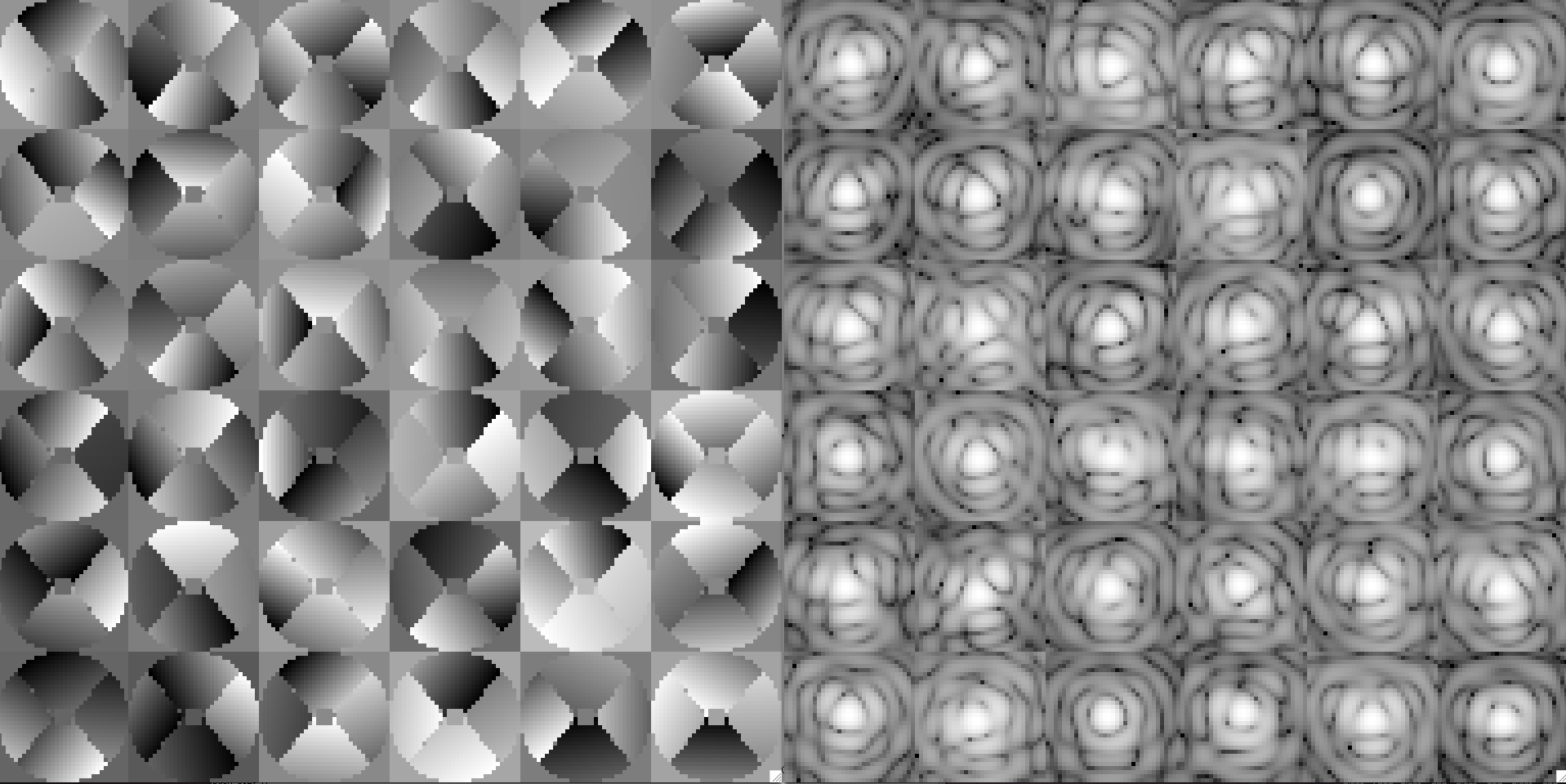}
	\end{tabular}
	\end{center}
   \caption[ex] 
   { 
Phase error (left) and corresponding DTTS PSF (right) for the simulation case assuming 400nm RMS of wind tip tilt, 100nm RMS of random tip tilt, 100nm RMS of piston tip tilt, and 20nm RMS of random low order aberrations.
   \label{fig_LWE_simulation} 
   }
   \end{figure} 

   \begin{figure} [ht]
   \begin{center}
   \begin{tabular}{c} 
   \includegraphics[width=0.6\hsize]{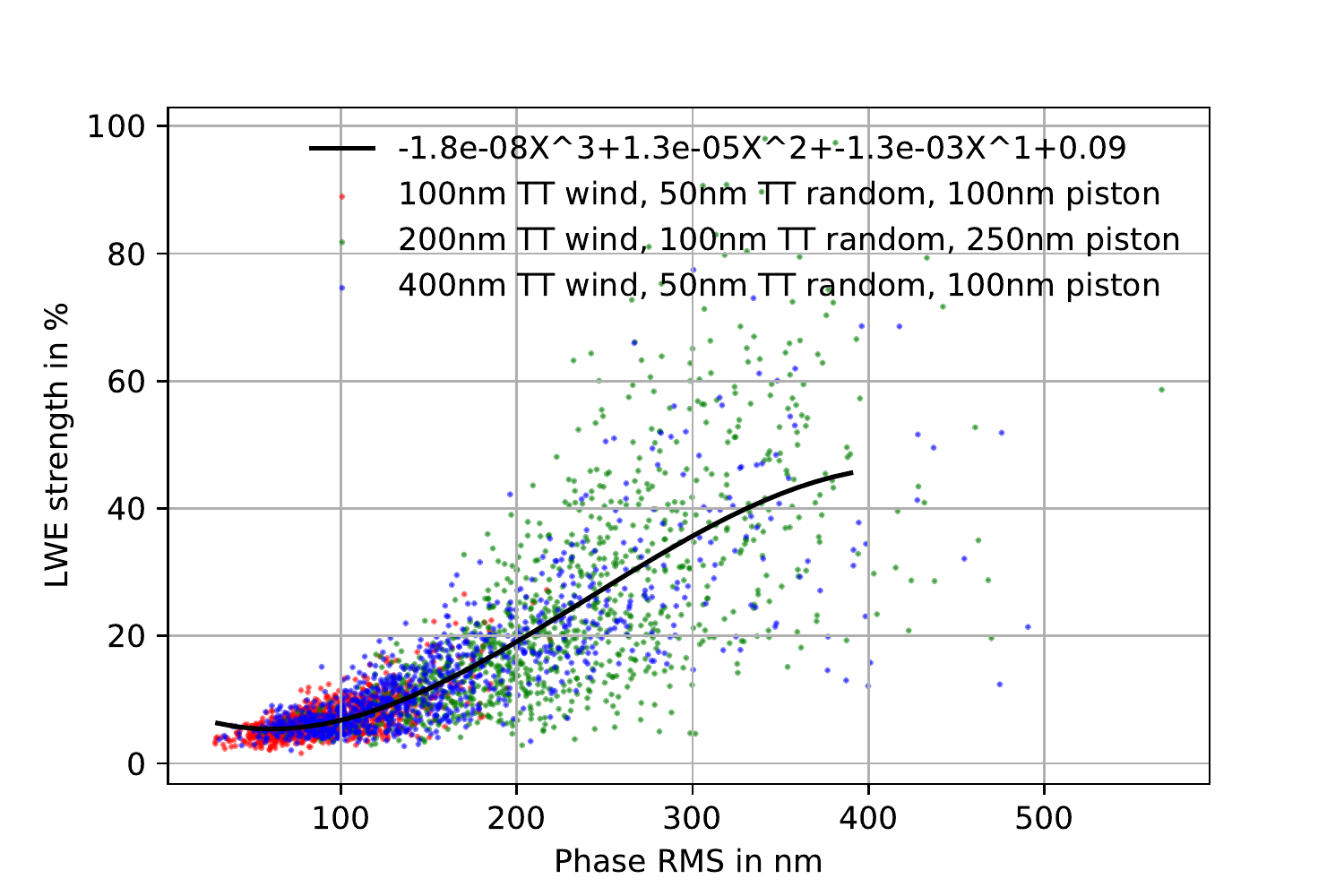}
	\end{tabular}
	\end{center}
   \caption[ex] 
   { 
Correspondence between the LWE strength and the phase error in nm RMS. The three colours correspond to the three scenarios used. A third degree polynomial was fitted to the point (plan black line).
   \label{fig_LWE_simulation_LWE_strength_vs_RMS_phase} 
   }
   \end{figure} 

We calibrated this parameter using simulated DTTS images. To do so, we produced synthetic DTTS images after introducing 3 types of differential piston and tip/tilt error on each pupil quadrant separated by the spiders:
\begin{itemize}
\itemsep0em 
\item a wind-induced tip-tilt  component whose value for each quadrant depends on an assumed direction of the wind. This distribution is not based on physical fluid dynamics results but only based on a pure geometrical approach.
\item a random tip-tilt (TT) on each quadrant. 
\item a random differential piston on each quadrant
\end{itemize}
On top of that, we assumed 20nm RMS of low-order aberrations, which is representative of the Non Common Path Aberrations seen in SPHERE\cite{Vigan2018} . 
 We then designed three test cases. For each test case, we made 1000 realisations of synthetic PSF, assuming that the RMS of each of the the three component described above (wind TT, random TT and random piston) follows a normal law of given standard deviation. The detailed amplitude of each phase error is shown in the legend of the graph of Fig. \ref{fig_LWE_simulation_LWE_strength_vs_RMS_phase}. 

We then measured the LWE strength on those synthetic images and plotted them as a function of the input phase error in Fig. \ref{fig_LWE_simulation_LWE_strength_vs_RMS_phase}. 
We see that a 20\% LWE strength corresponds to a phase error of about 200nm RMS, or a Strehl of about 60\%. Compared to the 80-90\% typical performance of SPHERE, this therefore represents a significant degradation of the performance by about 20-30\% Strehl. This confirms that the LWE strength can be used as a proxy for the phase error in the pupil of the telescope. A full Strehl measurement is not possible due to the low S/N of the DTTS image, especially for faint stars, and this parameter turns out to be a robust estimator.

We also calibrated this parameter by using DTTS images obtained with the internal lamp of SPHERE, in closed loop. We measured a LWE strength that could reach up to 7\%, depending on the level of low-order non common path aberrations present in the system. We therefore decided to use 10\% as a conservative threshold to claim the presence of LWE.

\subsection{Empirical dependance of the LWE strength $\mathcal{S}$}

Because the S/N can be small on the DTTS images, we associated an error with each measurement of $\mathcal{S}$, taking into account the readout and photon noise, and kept in the analysis values with S/N$>2$. We also removed manually a dozen binary targets for which the high value of $\mathcal{S}$ was due to the presence of a secondary companion near the first Airy ring. We ended up with
 46\,437 valid DTTS measurements or 8.1\% of the initial DTTS images.

   \begin{figure} [ht]
   \begin{center}
   \begin{tabular}{c} 
   \includegraphics[width=\hsize]{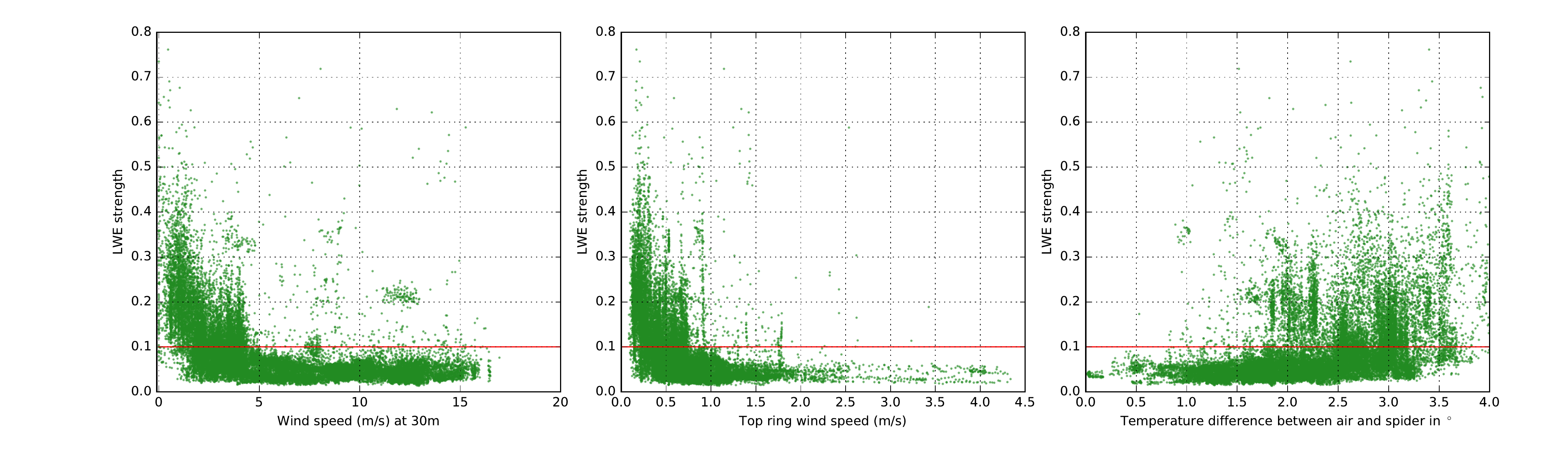}
	\end{tabular}
	\end{center}
   \caption[ex] 
   { 
LWE strength $\mathcal{S}$ as a function of the ambient wind speed (left), the top ring wind speed (middle) and the temperature difference between the spiders and the ambient air (right). The 10\% red line indicates the threshold above which the LWE can be robustly stated as present during the observations. 
   \label{fig_LWE_vs_windspeed_before_coating} 
   }
   \end{figure} 

\begin{figure} [ht]
   \begin{center}
   \begin{tabular}{c} 
   \includegraphics[width=0.8\hsize]{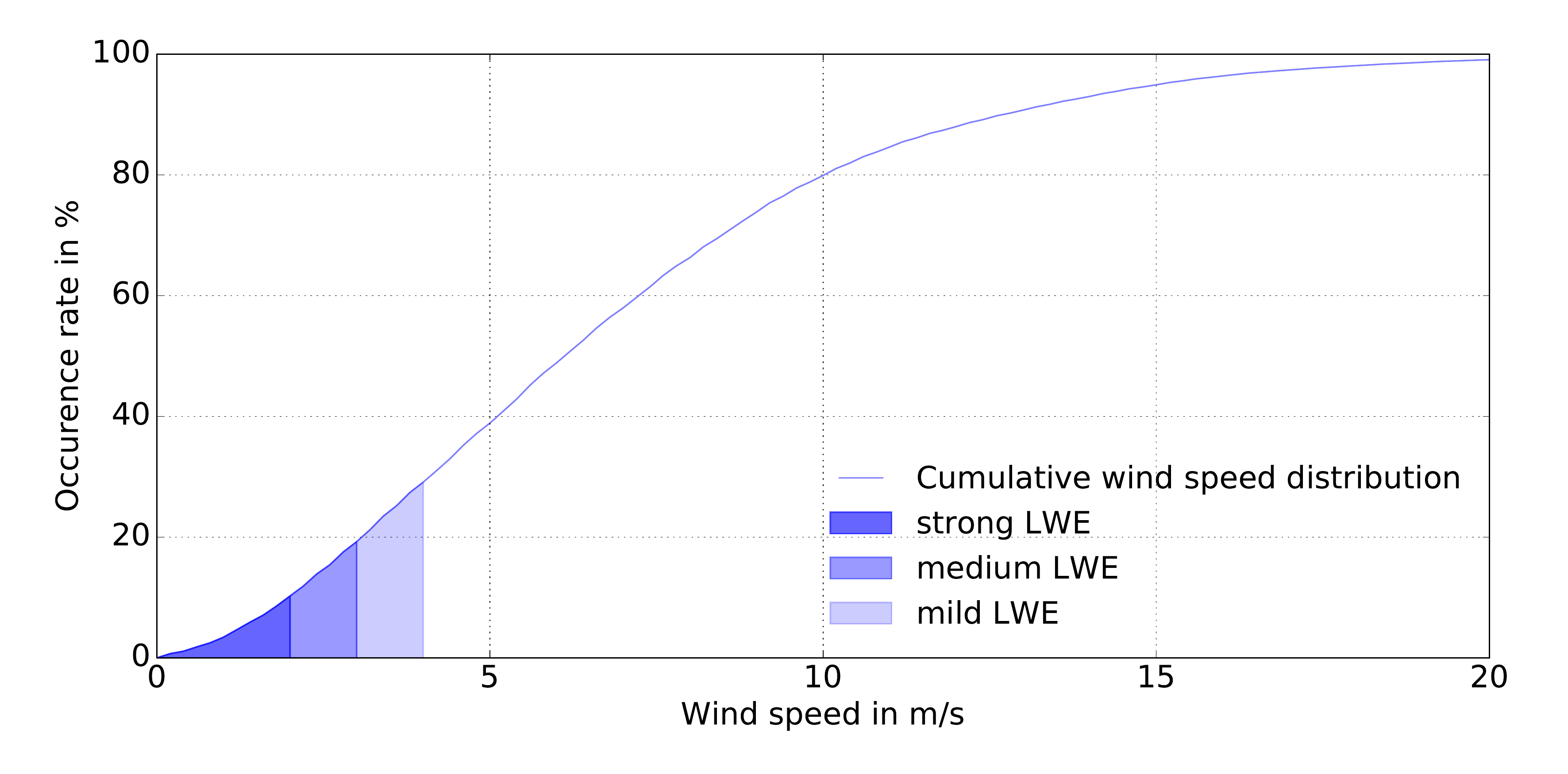}
	\end{tabular}
	\end{center}
   \caption[ex] 
   { 
Cumulative distribution function of the wind speed, as measured by the ASM (30m wind) from 2015 to 2018 during night time, highlighting the region when the LWE happens.
   \label{fig_windspeed_distribution} 
   }
   \end{figure}

Using the data from the start of the operations on January, 1 2015 up to the installation of a coating to mitigate the effect in August 2017, we computed the LWE strength as a function of the ambient wind speed measured by the meteorological tower at 30m. A clear trend is visible in Fig. \ref{fig_LWE_vs_windspeed_before_coating} (left). The LWE appears for a wind speed below 3m/s and the strength exceeds 20\% when the wind speed is below 2m/s. In terms of probability of occurrence, a wind below 4m/s represents 29\% of the Paranal night time as visible in Fig. \ref{fig_windspeed_distribution}, and it represented 15\% of the total time used by SPHERE observations from 2015 to 2018. Most importantly it happens more frequently in good conditions. If we define good conditions as a seeing below 0.8\arcsec and a coherence time above 4ms, then the wind is below 4m/s in 45\% of the time, which is therefore a significant issue for SPHERE. 

\begin{figure} [ht]
   \begin{center}
   \begin{tabular}{c} 
   \includegraphics[width=0.6\hsize]{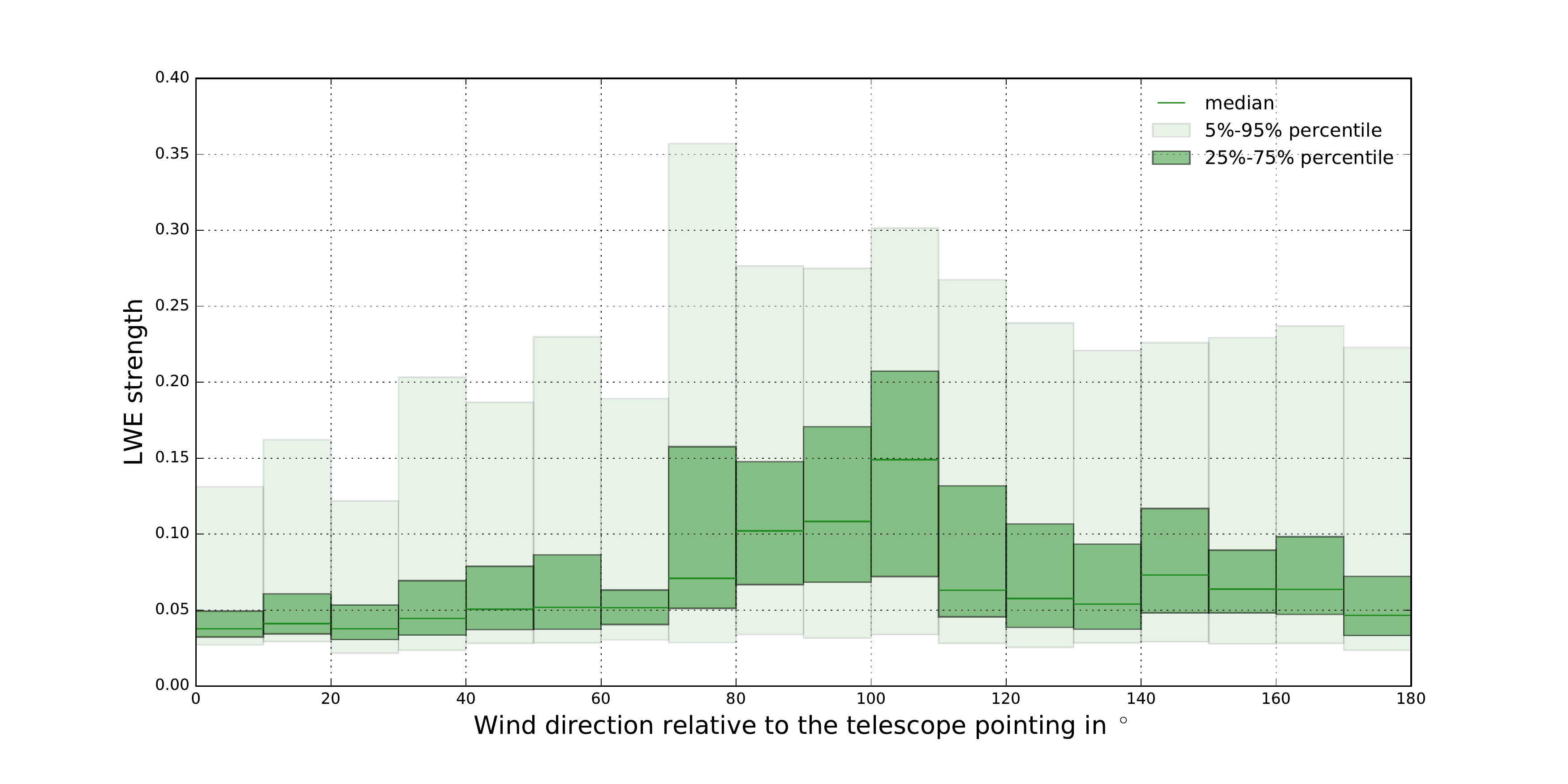}
	\end{tabular}
	\end{center}
   \caption[ex] 
   { 
Dependence of the LWE strength $\mathcal{S}$ as a function of the wind direction relative to the telescope pointing. A direction of $0^\circ$ means pointing to the wind while a direction of $180^\circ$ means pointing down the wind. 
   \label{fig_LWE_vs_winddir} 
   }
   \end{figure}

No dependence between the LWE strength and the telescope altitude could be concluded from the available data. An interesting result is the dependence of the LWE strength with the wind direction relative to the telescope pointing (Fig. \ref{fig_LWE_vs_winddir}). One would expect indeed that pointing to the wind yields the lowest LWE strength, as this increases air circulation within the dome, and this is indeed observed. A less intuitive observation is that the strongest LWE cases are not observed when the telescope is pointing exactly down the wind but at $90^\circ$ (either East or West). We think this is due the geometry of the dome and its lateral doors (see the picture in Fig. \ref{fig_zelda_phase_winddir} right) that block efficiently any wind coming perpendicularly, even when all lateral louvers are opened as it is the case in low wind conditions. The inadequate location of the anemometer on the top ring did not allow us to verify this assumption. We also investigated the dependance of the LWE strength with respect to the time ellapsed since the beginning of the night. The goal here was to detect whether the LWE needs a certain time before being able to settle. However the data did not show any dependance, the LWE was likely to occur at the beginning of the night than in the middle or in the end. 

\section{Corrective action}

\subsection{Application of a coating on the spiders}

A solution addressing directly the root cause of the problem at the telescope level was preferred over a solution at the instrument level, which would imply changes in the WFS strategy, whether it implies only software changes, such use of phase diversity algorithms with the DTTS\cite{Lamb2017, Wilby2018} or also hardware changes, such as the asymmetric Fourier pupil wavefront sensor \cite{Ndiaye2018}. The reason behind is that SPHERE is not the only VLT instrument sensitive to the LWE and the solution had therefore to be instrument-independent. 

\begin{figure} [ht]
   \begin{center}
   \begin{tabular}{c} 
   \includegraphics[width=0.8\hsize]{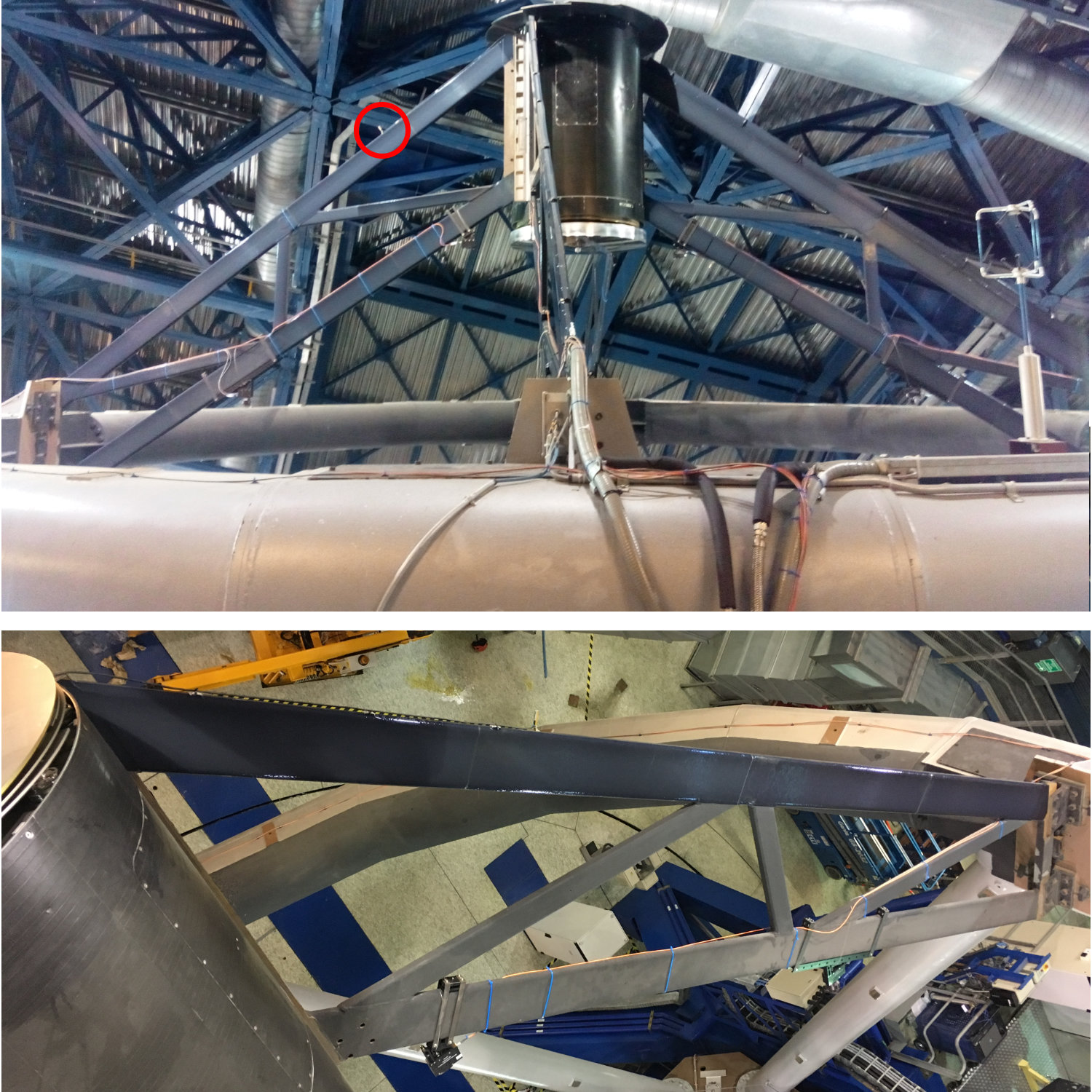}
	\end{tabular}
	\end{center}
   \caption[ex] 
   { 
Top: View of the top ring, secondary mirror, and spiders after the coating. The 3-axis supersonic anemometer of the top ring is also visible on the right and one of the two thermocouples used in the temperature analysis is shown with a red circle. Bottom: View of one of the four spiders while one of the two parallel 20cm-beams was coated and the other not.
   \label{picture_spiders} 
   }
   \end{figure}

\begin{figure} [ht]
   \begin{center}
   \begin{tabular}{c} 
   \includegraphics[width=\hsize]{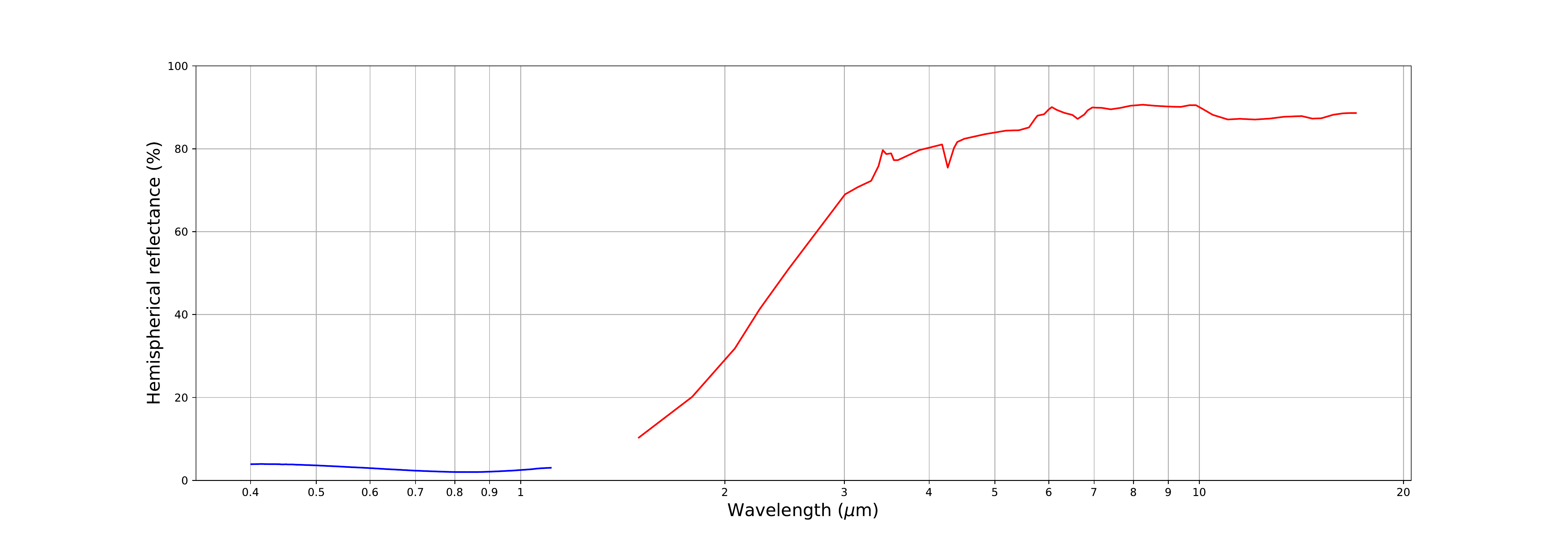}
	\end{tabular}
	\end{center}
   \caption[ex] 
   { 
Hemispherical spectral reflectance of the NanoBlack\textsuperscript{TM} foil as measured at optical and infrared wavelengths by the manufacturer, showing the sharp transition from absorbing to reflective material around 3\micron.
   \label{fig_nanoblack_properties} 
   }
   \end{figure}

Among the telescope-level solutions, one could think of active solutions such as actively warming the spiders or inducing forced airflow in the dome. The first one requires a redesign of the spiders and the second one is not recommended to avoid creating additional turbulence in the dome, detrimental to AO systems. Passive solutions could be designing optimised spider profiles or decrease the radiative transfer from the spiders to the sky by a coating, and the latter solution was implemented. A specific coating called NanoBlack\textsuperscript{TM} foil was bought from ACM Coatings GmbH, subsidiary company of Aktar Ltd, to cover the UT3 spiders. This foil is an aluminium-based substrate coated on one side with Acktar's NanoBlack\textsuperscript{TM} coating. The coating is applied in a proprietary vacuum deposition process in a roll-to-roll setup. It has a uniquely large specific surface area combined with very small coating thickness. The surface morphology (form and structure) is carefully controlled with dedicated measurements of physical, optical and radiation characteristics. 
The hemispherical reflectance as stated by the manufacturer is described in Fig. \ref{fig_nanoblack_properties}: it was measured at about 3\% in the range 450nm to 1100nm, with an emissivity of 13\%-14\% in the range 3\micron{} to 30\micron. The foil was applied to cover the lateral sides of the spiders (see Fig. \ref{picture_spiders}), namely: 
\begin{itemize}
\itemsep0em 
\item  the two 20cm-high parallel beams joining the M2 tower to the top ring
\item the three 10cm-high inner beams joining those 2 parallel bars
\end{itemize}

The coating of the four spiders was done between August, 8 2017 (coating of the first half of spider 1) and November, 19 2017 (coating of spider 4). Temperature data loggers were temporarily installed on the skin of one spider after the first coating in August 2017, and confirmed that a difference of $1.5^\circ$ between the uncoated side and the coated side, on average over 20 nights. 

In a second step, we systematically analysed the temperature difference between the air and the spiders before and after the four spiders of UT3 were coated. We used the two thermocouples installed on the northern and southern spider vanes. They are not in direct contact with the skin of the spiders, unlike the temperature data loggers described before, but give an estimation of the air temperature very near the spider. 

\begin{figure} [ht]
   \begin{center}
   \begin{tabular}{c} 
   \includegraphics[width=0.8\hsize]{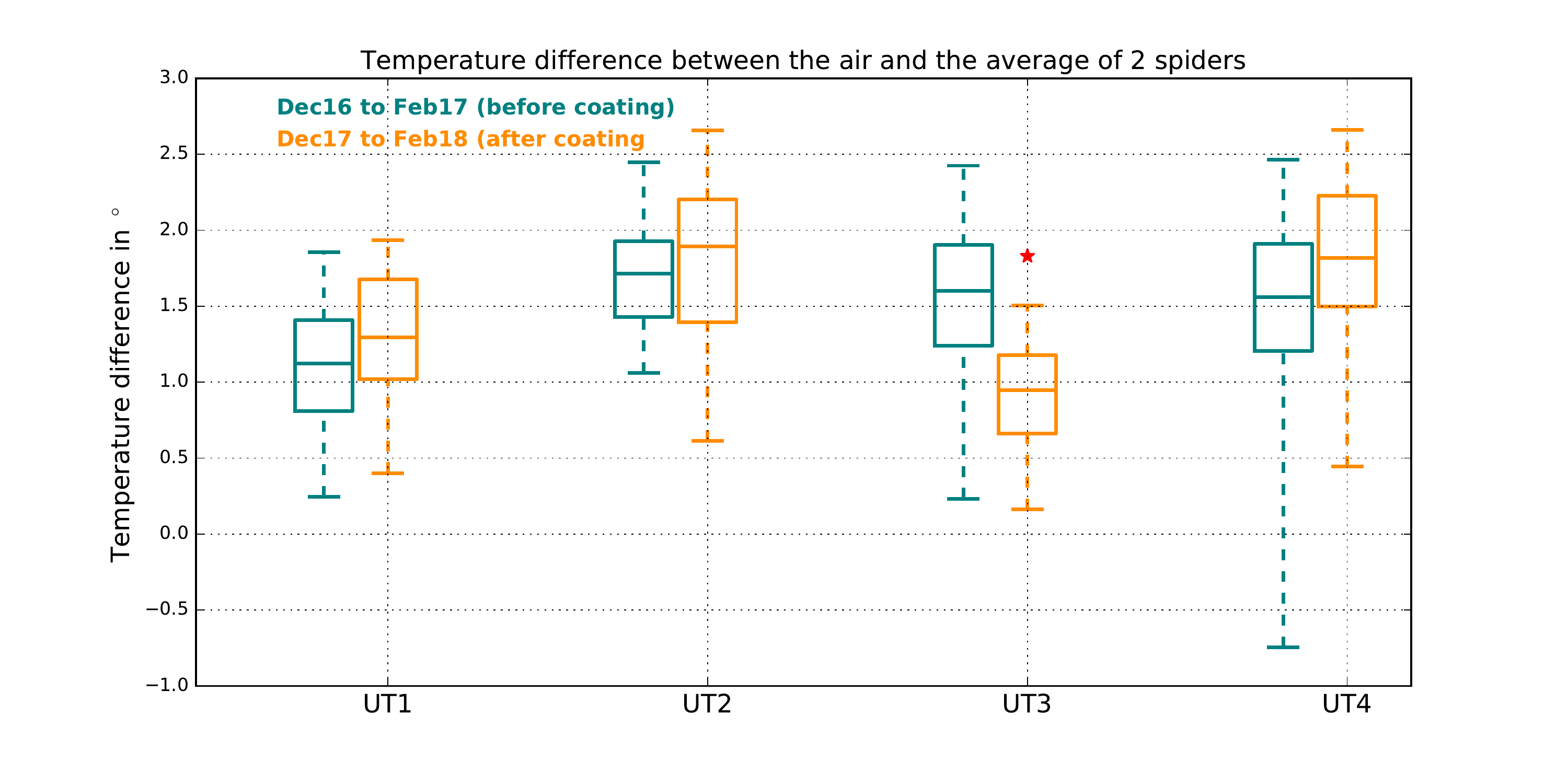}
	\end{tabular}
	\end{center}
   \caption[ex] 
   { 
Box-plot of the temperature difference between the air and the spiders for the 4 UTs and for 2 periods: before coating (green) and after coating (yellow). For each UT and period, the whisker plot indicates the 5\%, 25\%, 50\%, 75\% and 95\% percentiles. The red star for UT3 indicates where the median value would lie if it had followed the same evolution as the other UT between the two periods. Instead, the median temperature difference is 0.88º below. 
   \label{fig_whisker_plot_temperature_spiders} 
   }
   \end{figure}

Because these thermocouples have not been calibrated, an absolute comparison between them is not reliable. Therefore, we compared the temperature difference for each UT to data obtained one year before (same months) before the coating. These two epochs are shown in green and yellow colours in Fig. \ref{fig_whisker_plot_temperature_spiders}. 
It clearly shows that the thermal behaviour of UT3 spiders changed after the coating with respect to the other UTs. More quantitatively, we see that the temperature difference between the air and the spiders for UT1, UT2 and UT4 increased by 14\% on average between the 2 epochs (before coating and after coating), probably because of different ambient conditions between the 2 epochs. If UT3 had followed the same trend, the temperature difference would have been $1.83^\circ$ (red star in Fig. \ref{fig_whisker_plot_temperature_spiders}). Instead the temperature difference was limited to $0.95^\circ$, showing a decrease of $0.88^\circ$. In addition we can see that the temperature dispersion was also reduced for UT3 after the coating. 

\begin{figure} [ht]
   \begin{center}
   \begin{tabular}{c} 
   \includegraphics[width=0.8\hsize]{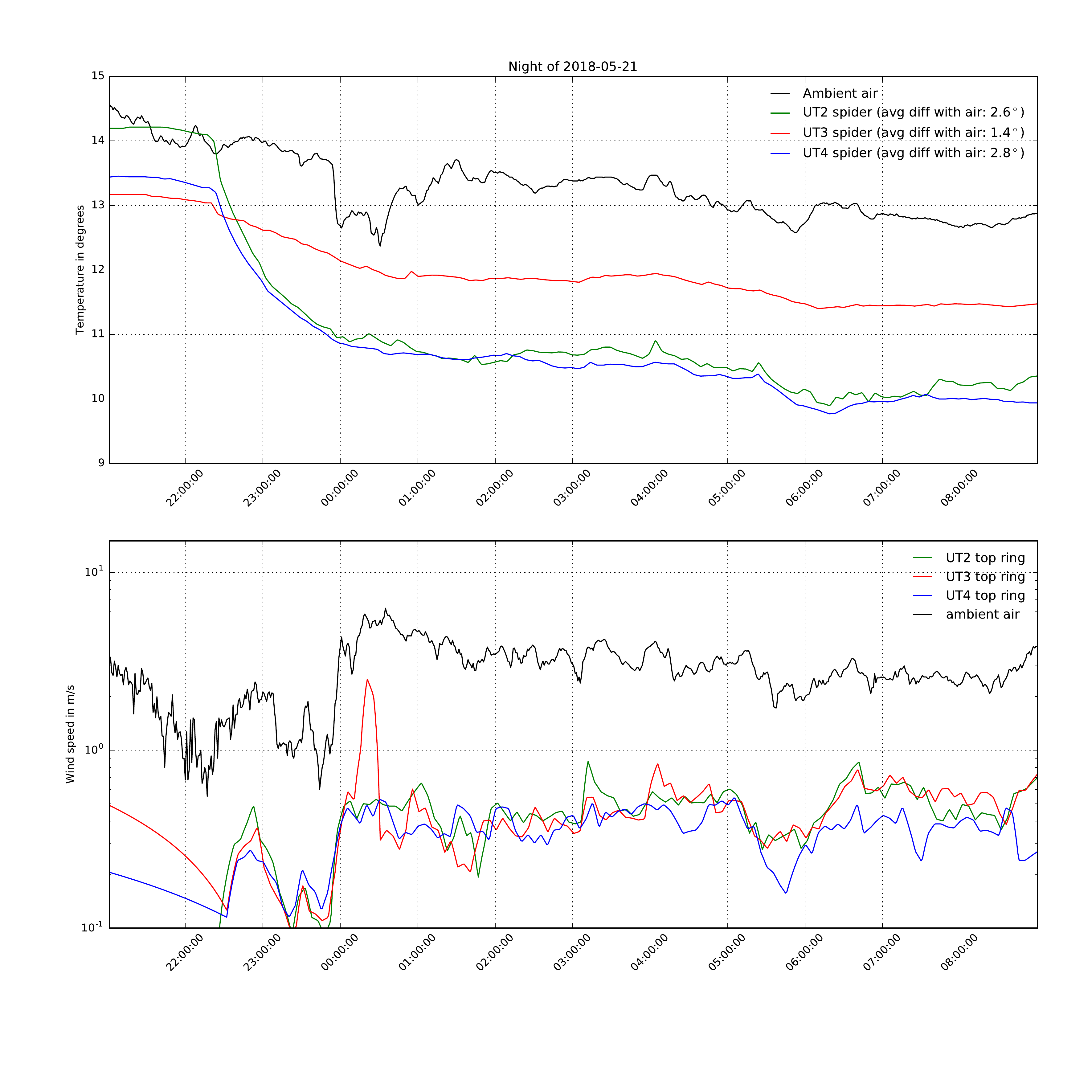}
	\end{tabular}
	\end{center}
   \caption[ex] 
   { Evolution of the temperature (top) and wind speed (bottom) during a night with low wind conditions (1m/s at the beginning and 3m/s afterwards). We purposely removed UT1 from the plot because one of the two thermocouples located on the spider vanes is out of order (we averaged the two sensor values for the other UTs) and the paint of UT1 is different.
   \label{fig_low_wind_night_spider_prop} 
   }
   \end{figure}

\begin{figure} [ht]
   \begin{center}
   \begin{tabular}{c} 
   \includegraphics[width=0.8\hsize]{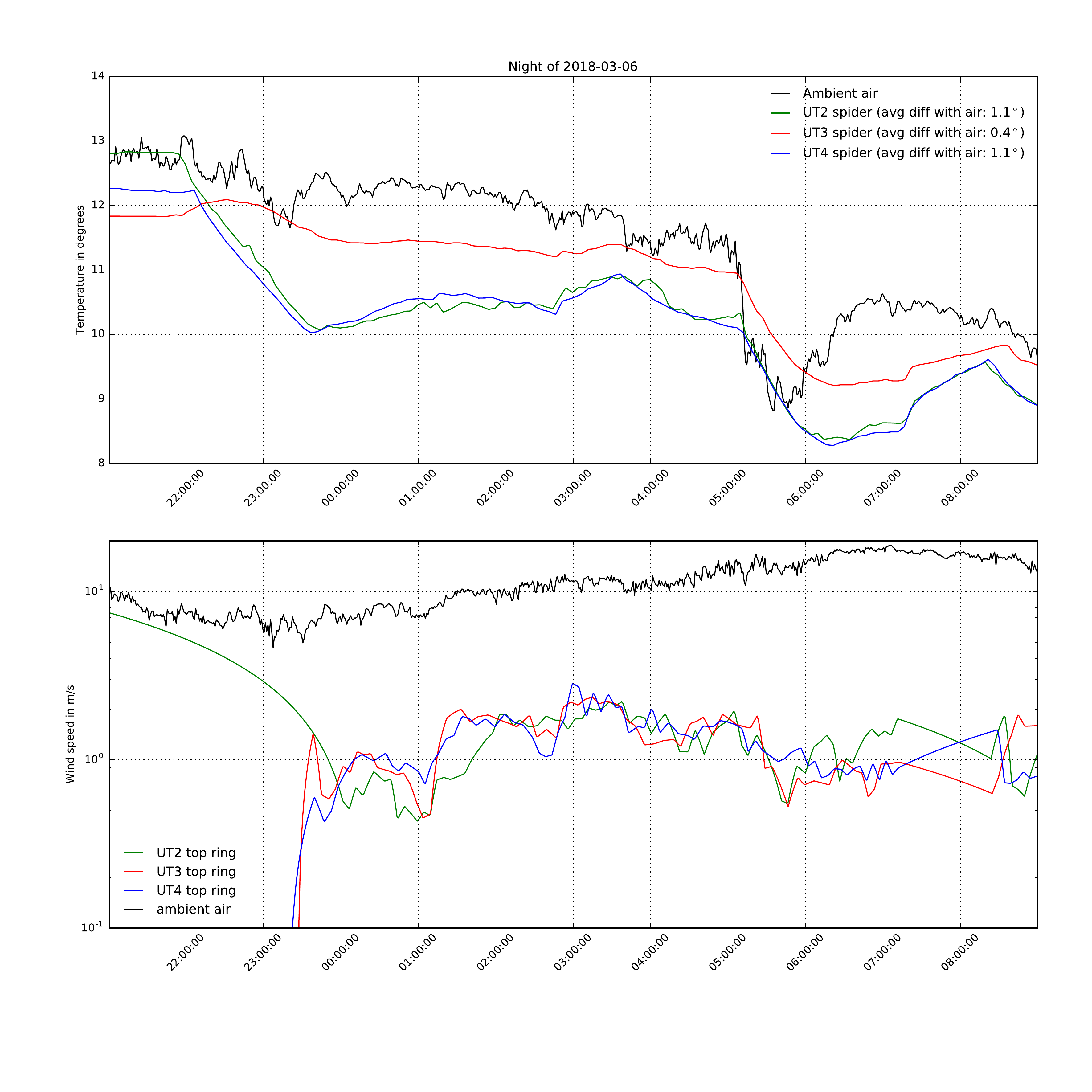}
	\end{tabular}
	\end{center}
   \caption[ex] 
   { Evolution of the temperature (top) and wind speed (bottom) during a night with moderate to high wind conditions.
   \label{fig_high_wind_night_spider_prop} 
   }
   \end{figure}

In addition to this global analysis, it is worth analysing the temporal evolution of the temperature of the spiders during the night, especially the comparison between the coated spiders of UT3 and the spiders of the other UTs. We show in Fig. \ref{fig_low_wind_night_spider_prop} and \ref{fig_high_wind_night_spider_prop} two nights in low wind conditions, and moderate to high wind conditions when all UTs were looking in the same direction to make sure the wind conditions were similar on the top ring of all UTs (VLTI nights). While the other UT's spiders cool down very quickly as soon as the dome is opened (around 22:00UT), the decrease is slower for UT3: the spiders of UT3 remain better coupled to the air temperature, with a mean difference twice smaller than the other UTs. We notice that the difference in the thermal behaviour happens all during the first two hours of the night. After that, an equilibrium is reached between the heat loss by radiation to the sky and the heat gain by convection with the warmer ambient air and the temperature evolution of the spiders follows that of the air, with an offset different for UT3 and for the other UTs.

\subsection{Impact of the coating on the LWE}

\begin{figure} [ht]
   \begin{center}
   \begin{tabular}{c} 
   \includegraphics[width=\hsize]{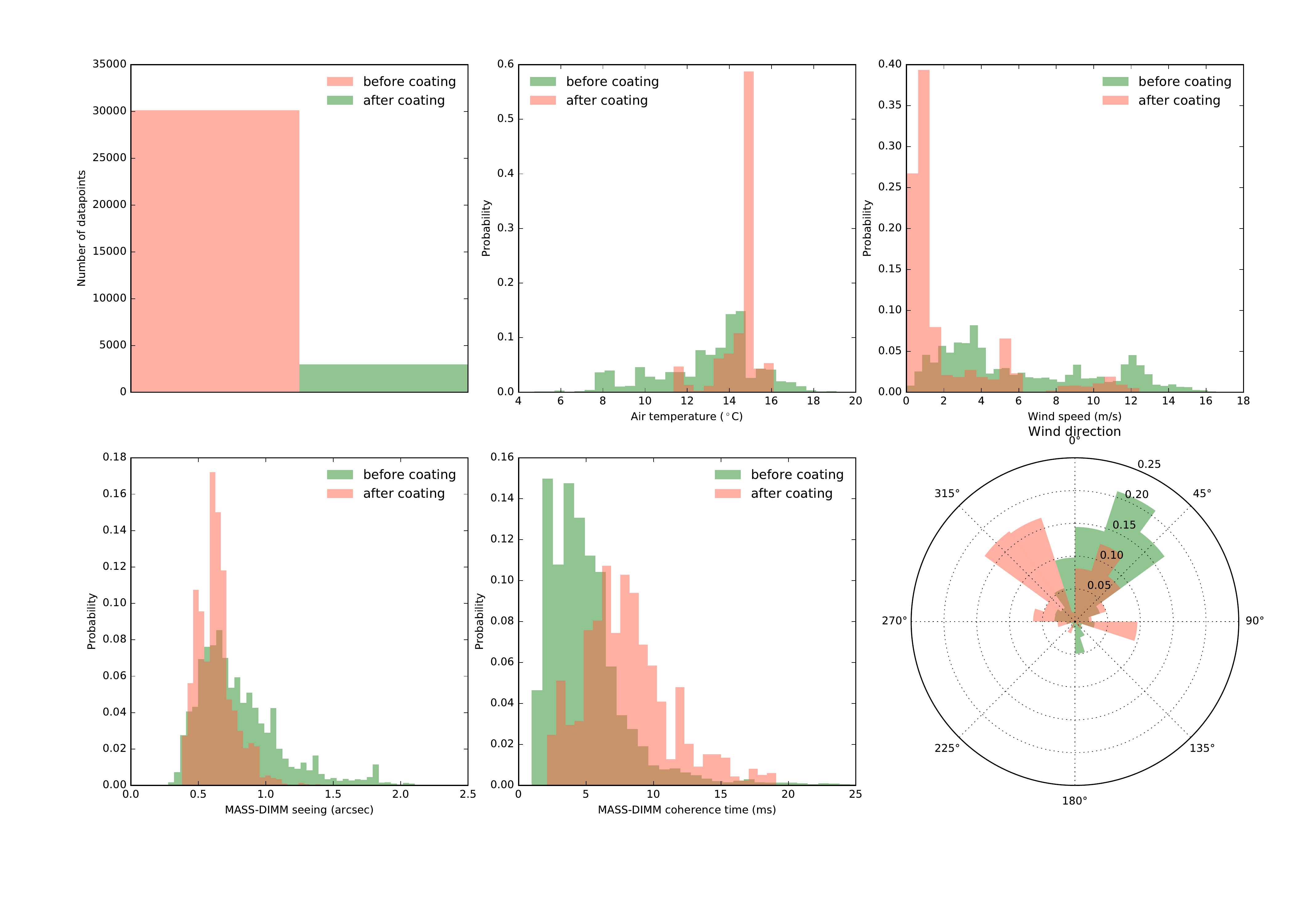}
	\end{tabular}
	\end{center}
   \caption[ex] 
   { 
Histograms of the ambient and turbulence conditions corresponding to the valid LWE strength data used in the analysis. There is ten times more valid data before the coating than after (top right histogram).
   \label{fig_summary_conditions_before_after_coating} 
   }
   \end{figure}

We have currently about six months of statistics after the coating was applied, corresponding mostly to the summer season in Paranal where conditions are more favourable in terms of turbulence. We compare in Fig \ref{fig_summary_conditions_before_after_coating} the ambient and turbulence conditions for the data for which the LWE strength passed the S/N threshold. It shows that there were many low wind nights, which is good as this is the best environment to test the efficiency of our mitigation strategy.

\begin{figure} [ht]
   \begin{center}
   \begin{tabular}{c} 
   \includegraphics[width=\hsize]{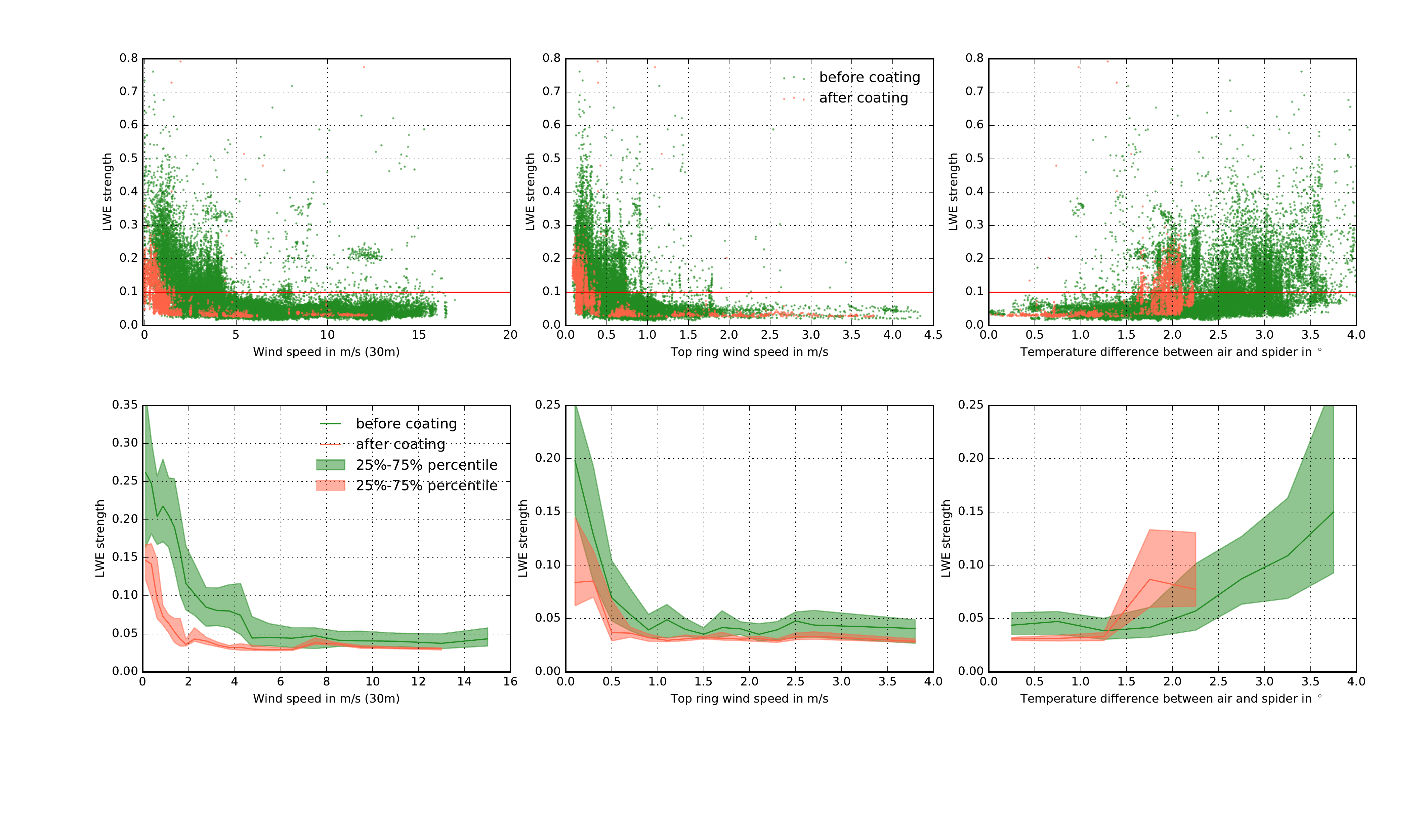}
	\end{tabular}
	\end{center}
   \caption[ex] 
   { 
Same plot as in Fig. \ref{fig_LWE_vs_windspeed_before_coating} comparing the LWE strength before coating the spiders (green) and after (orange). The bottom graphs are the same as the top ones binned to highlight the impact of the coating. The plain line is the median while filled areas indicate values between the 25\% and 75\% percentiles.
   \label{fig_LWE_vs_windspeed_after_before_coating} 
   }
   \end{figure}

\begin{figure} [ht]
   \begin{center}
   \begin{tabular}{c} 
   \includegraphics[width=\hsize]{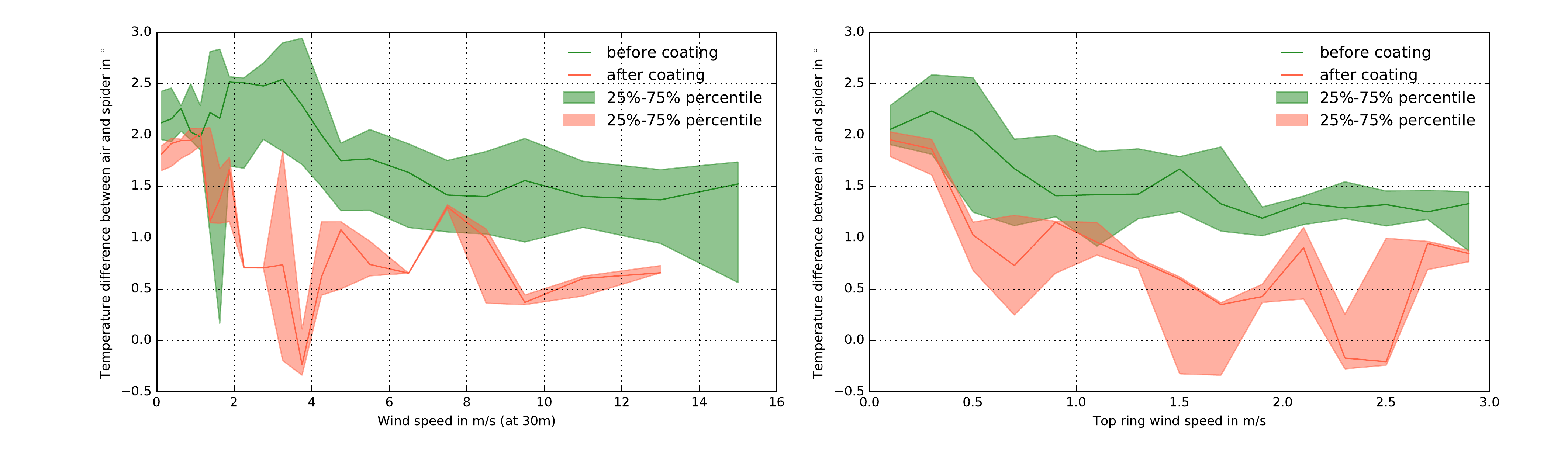}
	\end{tabular}
	\end{center}
   \caption[ex] 
   { 
   Correlation between the temperature difference between the air and the spiders, and the wind speed, as measured by the meteo tower (left) or at the top ring level (right). 
   \label{fig_correlation_temp_diff_vs_windspeed} 
   }
   \end{figure}

The analysis of the LWE strength after the coating shows that the effect was significantly decreased but not totally removed: it appears now only in conditions of wind below $\sim1$m/s and its strength is significantly reduced, with no more PSF completely split without a coherent core. The statistics is shown in Fig. \ref{fig_LWE_vs_windspeed_after_before_coating}. The wind speed as measured by the top ring anemometer needs to be below 0.5m/s for the LWE to appear. There is no more temperature difference greater than $2.5^\circ$ between the air and the spiders, as visible in Fig. \ref{fig_LWE_vs_windspeed_after_before_coating} right. This graph might give the impression that the LWE is now stronger for a temperature difference between $1.5^\circ$ and $2^\circ$. But much smaller windspeed are now required to produce such a temperature change, meaning that a given volume of air stays longer in contact with the colder spiders. This correlation between temperature difference and wind speed is given shown in more details in Fig. \ref{fig_correlation_temp_diff_vs_windspeed}.

\begin{figure} [ht]
   \begin{center}
   \begin{tabular}{c} 
   \includegraphics[width=\hsize]{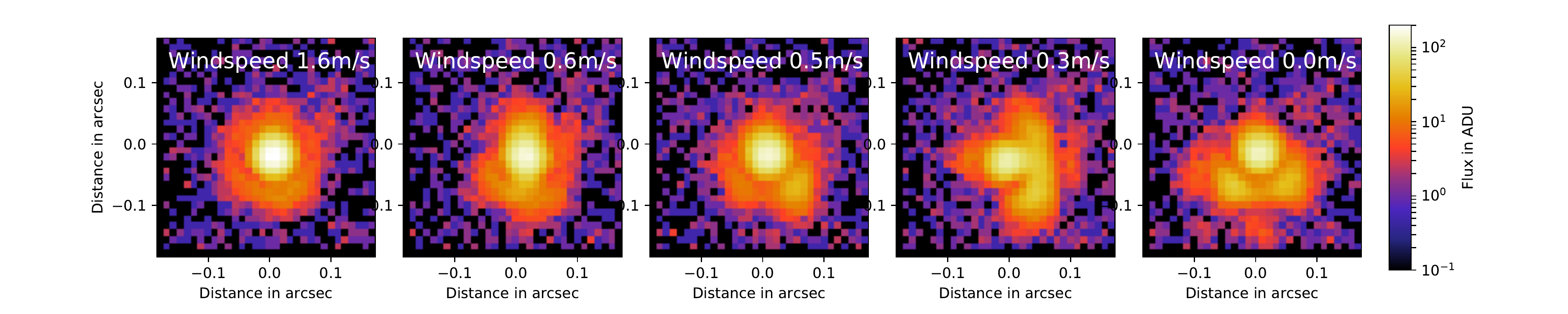}
	\end{tabular}
	\end{center}
   \caption[example] 
   { \label{fig_LWE_video} 
Sequence of DTTS images obtained while the wind dropped from 2m/s to almost 0. The secondary lobes typical from the LWE appeared once the wind had dropped below 1m/s. A video is available online at \url{https://doi.org/10.1117/12.2311499} (file dtts\_movie\_2018-01-04.mov in the supplemental content section).} 
   \end{figure}

A convincing example showing that the LWE still happens below 1m/s is shown in the video of Fig. \ref{fig_LWE_video}. No effect is detectable above 1m/s. If we believe this new threshold, this means the probability of occurrence is decreased from 19.4\% of the time (wind below 3m/s) down to 3.5\% of the time (wind below 1m/s) as shown in the cumulative distribution of the wind speed in Fig. \ref{fig_windspeed_distribution}. From the operations point of view, this fraction of the time becomes now manageable, in the sense that a change of focus towards instruments less sensitive to this effect (VISIR, ESPRESSO) can be done in the 3.5\% of the time when a mild LWE is detectable on SPHERE and the probability and impact is now low enough not to penalise significantly visitor observers.  However, the fact that it impacts the best nights in terms of turbulence especially coherence time still represents an unsolved issue, considering that coherence time will become in 2019 a user constraint for SPHERE observations. Software-only solutions\cite{Wilby2018} exist to tackle this remaining effect or a new pyramid WFS sensitive to the  LWE might be introduced as part of a future upgrade of the instrument.

\section{LWE on other VLT instruments and other telescopes}

\subsection{Impact on other VLT instruments}

\begin{figure} [ht]
   \begin{center}
   \begin{tabular}{c} 
   \includegraphics[width=\hsize]{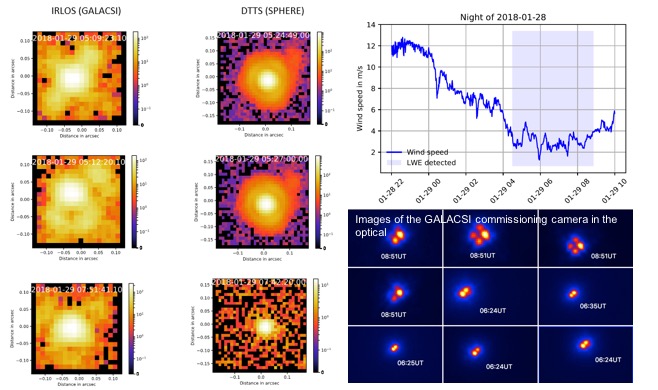}
	\end{tabular}
	\end{center}
   \caption[example] 
   { \label{fig_LWE_galacsi} 
 Comparison between AO-corrected images obtained at UT3 with SPHERE (DTTS images in the middle) and at UT4 (IRLOS camera in the near-infrared similar to the DTTS and GALACSI optical camera on the bottom right corner) during the low wind night of January 28 2018. 
A video of IRLOS is available online at \url{https://doi.org/10.1117/12.2311499} (file video\_IRLOS\_2018-01-28\_LTAO\_nphot0.5.mov in the supplemental content section).
}
   \end{figure}

SPHERE is not the only AO-instrument of the VLT where the LWE is noticeable. The clearest sign of the LWE was seen during the recent commissioning of the Laser Tomography Mode of MUSE, also known as MUSE Narrow Field Mode (NFM)\cite{Oberti2016}. Interestingly, SPHERE on UT3 and MUSE NFM on UT4 were observing simultaneously on the commissioning night of January 28 2018 when the wind was below 4\,m/s. Fig. \ref{fig_LWE_galacsi} clearly shows that the LWE is not detected on UT3 on that night (DTTS image shown in the middle) whereas it is clearly present on UT4 (IRLOS images shown on the left, IRLOS is the near-infrared tip-tilt sensor of GALACSI, MUSE AO system) at similar timestamps. At this time, the wind speed was between 2 and 4m/s as shown by blue shaded area in the top right plot of Fig. \ref{fig_LWE_galacsi}. This confirms the previous conclusion that the threshold when the LWE occurs decreased due to the coating of UT3, from about 3-4m/s to 1m/s. The distortion seen on UT4 is worse in the optical wavelength range, as shown in Fig. \ref{fig_LWE_galacsi} bottom left. The core of the PSF is split, as also noticed on SPHERE for the worst conditions. For this reason, the spiders of UT4 are currently being coated with the same low-emissivity material as the one used for UT3, to allow nominal performance of MUSE NFM even under low wind conditions. 

The VLT-interferometer (VLTI) also hosts a new infrared WFS called CIAO\cite{Kendrew2012} that was interfaced with the MACAO deformable mirrors and delivers between 10 and 35\% Strehl in the K band. So far, no systematic analysis of the LWE strength during VLTI runs with the four UTs has been carried out, but we performed a preliminary analysis on a single night. For that we selected the VLTI run with the least wind during a run on March, 31 2018. It indicates that the PSF is more symmetric on UT3 than on the other UTs. The measurement of the LWE strength $\mathcal{S}$ indicates a value of $\mathcal{S}_3=4.9\%\pm1.6\%$ for UT3, whereas the values were higher for the other UTs $\mathcal{S}_1=5.4\%\pm1.7\%$, $\mathcal{S}_2=7.5\%\pm2.5\%$ and $\mathcal{S}_4=9.8\%\pm3.9\%$ ($1\sigma$ error bars). 

On systems equipped with AO systems of the earlier generation such as NaCo (UT1) and SINFONI (UT4), the LWE effect was also present but clear evidence of the effect is harder to demonstrate as the Strehl level is not as high as for SPHERE and other problems such as miscalibrations can produce similar effects. The variation in brightness of the spider diffraction shown in Fig. \ref{fig_spider_diffraction_spike} spikes was clearly  visible with NaCo for instance. 
With the mid-infrared imager VISIR on UT3, there are also hints of LWE in some short-exposure PSFs, as the instrument is diffraction-limited in the Q band, and also in the N band under good seeing conditions\cite{Vandenancker2016} . However due to other sources of image elongation independent of the LWE, tracing the root cause of the effect is here again not straight-forward.


\subsection{Impact on other high-contrast instruments at other observatories}

\begin{table}[ht]
\caption{ Geometrical and thermal properties of some telescope spiders on large telescopes with AO systems and high-contrast instruments} 
\label{tab_other_telescopes}
\begin{center}       
\begin{tabular}{| l | l | l | l | l | l | } 
\hline
\rule[-1ex]{0pt}{3.5ex}  Telescope & VLT UT3 & Gemini & Magellan II & Subaru  & Keck II\\
\rule[-1ex]{0pt}{3.5ex}   &  & South & (Clay) &   & \\
\hline
\rule[-1ex]{0pt}{3.5ex}  High-contrast  & SPHERE & GPI   & MagAO & CHARIS	 & NIRC2 \\
\rule[-1ex]{0pt}{3.5ex}  instrument &  &    &  & Vampires	 &  \\
\hline
\rule[-1ex]{0pt}{3.5ex}  LWE occurence &  3\% now & $<3\%$$^a$   & Never & $\sim$10\%-20\%$^a$ & Never \\
\rule[-1ex]{0pt}{3.5ex}    &  (20\% before) &     &   &   &   \\
\hline
\rule[-1ex]{0pt}{3.5ex}  WFS & 40x40 SH & 43x43 SH  & 28x28 Py & 50x50 Py & 21x21 SH \\
\hline
\rule[-1ex]{0pt}{3.5ex}  Spider thickness & 5cm & 1cm &   3.8cm & 23cm & 2.5cm \\
\hline
\rule[-1ex]{0pt}{3.5ex}  Ratio spider thickness  & 25\% & 5\% & 16.5\% & 145\% & 4.4\%  \\
\rule[-1ex]{0pt}{3.5ex}  over subpupil  &  & & & & \\
\hline
\rule[-1ex]{0pt}{3.5ex}  Spider height & 20cm &10cm & 12.7cm & up to $\sim$1m &15cm \\
\hline 
\rule[-1ex]{0pt}{3.5ex}  Spider coating & NanoBlack\textsuperscript{TM} & Chemglaze$^b$   & Cadmium  &NA & low-em.paint  \\
\rule[-1ex]{0pt}{3.5ex}   & (Chemglaze$^b$ before) &    &  plating & &  ($\epsilon<0.25$)\\
\hline 
\end{tabular}
\end{center}
{Notes. $^a$ Estimation. $^b$ Chemglaze is also known as Aeroglaze Z306 polyurethane flat black absorptive paint}
\end{table}

   \begin{figure} [ht]
   \begin{center}
   \begin{tabular}{c} 
   \includegraphics[width=0.5\hsize]{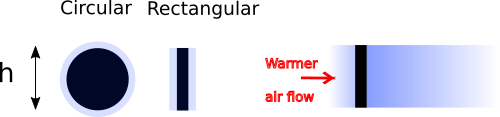}
	\end{tabular}
	\end{center}
   \caption[example] 
   { \label{fig_spider_drawing} 
   Comparison between a circular cross-section and a rectangular cross-section in the presence of  a surface layer of cold air (blue shaded area) in contact with the spider skin. On the right, we show the asymmetric temperature distribution that results from a slow laminar flow and that can lead to a phase map as shown in Fig. \ref{fig_zelda_phase_winddir}.
}   \end{figure} 

The presence of LWE depends on the cross-section of the spiders, their dimension, their thermal properties and the ability of the WFS to see the aberration and of the deformable mirror to correct for it. We summarised in Table \ref{tab_other_telescopes} those parameters for five telescopes equipped with high-contrast imagers for which we could obtain some data (see acknowledgements section). 

For the same height, a rectangular cross-section is more likely to induce wavefront errors than a circular one, as it was already noticed by Couder in 1949\cite{Couder1949}. In the presence of mild wind, if we assume the colder spider is surrounded by a surface layer of air colder than the ambient air, the path travelled by the light in this colder surface layer is indeed smaller for a circular section than for a rectangular section (Fig. \ref{fig_spider_drawing} left) and phase maps obtained with Zelda with mild wind confirm the presence of this surface layer around the VLT spiders of thickness a few centimetres. In the presence of very low wind or absence of wind, a wall constitute an obstacle much more difficult to overcome than a cylinder, and this absence of convection downstream of the spider can create the type of phase maps shown in Fig. \ref{fig_zelda_phase_winddir}, sketched in Fig. \ref{fig_spider_drawing} right, and supported by the Computer Fluid Dynamics simulations of Sauvage et al. 2016\cite{Sauvage2016} .

Obviously, the higher the height of the spider, the larger the surface area exposed to the cold sky and radiating heat away, and the more efficient the heat removal from the ambient air around the spider. This parameter seems too be the critical part in the case of the Subaru telescope, equipped with very high monolithic spiders compared to the other telescopes that have several beams but with a smaller height. 

For a hollow tube, the thickness of the spider as projected on the pupil probably does not play a significant role in the heat transfer, but it has an effect on the detectability of the phenomenon with the WFS. While extreme AO systems have now very fine WFS pupil sampling, the fraction of the sub-pupil obstructed by the spiders becomes significant, up to 25\% in the case of SPHERE that has 5cm-thick spiders for sub-pupils of 20cm. Sauvage et al. 2016\cite{Sauvage2016} showed that the variance of the residual slopes is indeed significantly larger in sub-pupils obstructed by the spiders.  

Last, the coating or paint of the telescope spiders helps to reduce the effect as it was demonstrated in this study, and the two other telescopes equipped originally with low-emissivity paint, Magellan II and Keck II, do not suffer from the LWE.

\section{Conclusions}

The LWE strength was assessed quantitatively using the DTTS images on SPHERE and it was shown that the effect started to appear when the ambient wind speed dropped below 3 to 4\,m/s. A strong effect appeared below 2\,m/s. It significantly degrades the PSF, with a relative decrease in Strehl by 20\% to 30\% for a LWE strength of 20\%. To mitigate this effect, a specific coating was applied to the spiders of UT3, which characteristics are a low emissivity in the thermal infrared above 3\micron{} and a low reflectance in the optical below 1\micron. This coating was fully applied to the 4 spiders after November, 19 2017. The effect of the coating on the spider temperature is clearly demonstrated: the UT3 spiders skin temperature is $1.5^\circ$ higher with the coating, and on average the air temperature around the spiders is $0.9^\circ$ higher, which was the gain expected with the coating from a thermal analysis performed internally at ESO. This change in the thermal behaviour of the spider is directly reflected in the image quality as well: for wind conditions between 1 and 3m/s, the LWE is no longer detectable on the DTTS images. This means the probability of occurrence of the effect was decreased from 20\% to 3.4\% thanks to the coating. This demonstrates the efficiency of the new coating to mitigate strongly the LWE. Nonetheless, it did not completely remove the effect: from December 2017 to May 2018, medium-strength LWE events occurred when the wind speed was below 1\,m/s. Several solutions based on modifications of the AO control loop and wavefront sensing strategy have already been proposed for SPHERE\cite{Lamb2017, Wilby2018,Ndiaye2018} or other high-contrast instruments, and might be implemented in a future upgrade of the instrument\cite{Beuzit2018} . 

The LWE was also clearly detected on UT4 at the VLT, both in the near-infrared and at optical wavelengths. Therefore, the same coating is being applied on UT4. Two other aspects must be carefully looked at now. On the one hand, the ageing of the coating must be monitored. On the other hand, the impact of the higher specular reflectance of the NanoBlack\textsuperscript{TM} foil needs to be considered for instruments working at optical wavelengths, such as MUSE . An analysis of the stray light, background level and possible additional laser light scattered through the Raman effect\cite{Vogt2017} is currently on-going for UT4. At other observatories with 8m-class telescopes and high-contrast imagers, the occurence of LWE varies from inexistent (Magellan, Keck) to rare (Gemini South) or frequent (Subaru), depending on the spider shape, dimension and thermal properties, indicating that optimal spider designs to avoid this harmful effect exist and need to be considered for any new facility of that class. For the new class of giant segmented telescopes currently being built, reaching the diffraction limit at optical or near-infrared wavelengths is among the science requirements of most of their instruments to take the full benefit of the size of the primary mirror. In this condition, the LWE must be carefully taken into account in addition to the more general island effect that can be generated by the segments of the primary mirror. The spiders will be much thicker and higher to support a four-meter class secondary mirror weighing 3.5 tons in the case of the European Large Telescope. On the one hand, an increased surface area will lead to higher radiation loss to the night sky, but on the other hand, massive plain spiders will have a much larger heat capacity so that the fast temperature experienced by the spiders right after the dome opening might be significantly delayed. These considerations will be critical to select the best solution to address this issue on extremely large telescopes, and decide between active or passive mitigation strategies.

%
%
%

\acknowledgments 
 
J.M. thanks ESO staff and technical operators at the Paranal Observatory, especially the System \& Optics team for their work to apply the low-emissivity coating on the telescope spiders, and the Software team to allow to use an early release of the ESO Data Lab\cite{Pena2018} for the statistical analysis presented in this paper. He thanks W. Brandner for providing him VLTI data during a night with low wind, to evaluate the symmetry of the PSF on all 4 UTs. He thanks V. Bailey, B. Macintosh, G. Perez, M. Boccas, P. Wizinowich, J. Males, L. Close, J. Lozi and S. Vievard for providing information on the LWE and spiders properties on Gemini South, Magellan, Keck and Subaru telescopes respectively. He thanks P. Figueira for interesting discussion on statistical estimators of the LWE occurence rate. Last but not least, he thanks J. Smoker for his careful english language editing. This research has made use of the following python packages: matplotlib, pandas, scipy and astropy.

\bibliography{report} 
\bibliographystyle{spiebib} 

\end{document}